\documentclass[]{emulateapj}

\shorttitle{HOD Modeling of LRG Clustering}
\shortauthors{Zheng, Zehavi, Eisenstein, Weinberg, \& Jing}

\def\NavgM{\langle N(M)\rangle}
\def\Ncen{\langle N_{\rm cen}(M)\rangle}
\def\Nsat{\langle N_{\rm sat}(M)\rangle}
\def\Mmin{M_{\rm min}}
\def\sigM{\sigma_{\log M}}

\def\hMsun{h^{-1}M_\odot}
\def\erf{{\rm erf}}
\def\hMpc{h^{-1}{\rm Mpc}}

\begin{document}

\title{Halo Occupation Distribution Modeling of Clustering of Luminous Red 
Galaxies}
\author{
Zheng Zheng\altaffilmark{1,2,3}, 
Idit Zehavi\altaffilmark{4}, 
Daniel J. Eisenstein\altaffilmark{5}, 
David H. Weinberg\altaffilmark{6}, 
and Y.P. Jing\altaffilmark{7}
}
\altaffiltext{1}{
 Institute for Advanced Study, Einstein Drive, Princeton, NJ 08540;
 zhengz@ias.edu
}
\altaffiltext{2}{
 Hubble Fellow
}
\altaffiltext{3}{
 John Bahcall Fellow
}
\altaffiltext{4}{
  Department of Astronomy, Case Western Reserve University,
  10900 Euclid Avenue, Cleveland, OH 44106; izehavi@astronomy.case.edu
}
\altaffiltext{5}{
  Steward Observatory, University of Arizona, 933 N. Cherry Avenue, 
  Tucson, AZ 85121; deisenstein@as.arizona.edu
}
\altaffiltext{6}{
  Department of Astronomy and Center for Cosmology and AstroParticle Physics, 
  Ohio State University, 140 W. 18th Avenue, 
  Columbus, OH 43210; dhw@astronomy.ohio-state.edu
}
\altaffiltext{7}{
  Shanghai Astronomical Observatory, Joint Institute for Galaxy and Cosmology
  (JOINGC) of SHAO and USTC, Nandan Road 80, Shanghai, 200030, China; 
  ypjing@shao.ac.cn
}

\begin{abstract}
We perform Halo Occupation Distribution (HOD) modeling to interpret
small-scale and intermediate-scale clustering of 35,000 luminous
early-type galaxies and their cross-correlation with a reference imaging
sample of normal $L_*$ galaxies in the Sloan Digital Sky Survey. 
The modeling results show that most of these luminous red galaxies (LRGs) 
are central galaxies residing in massive halos of typical mass $M \sim$ 
a few times $10^{13}$ to $10^{14} \hMsun$, while a few percent of them 
have to be satellites within halos in order to produce the strong 
auto-correlations exhibited on smaller scales. The mean luminosity $L_c$ 
of central LRGs increases with the host halo mass, with a rough scaling 
relation of $L_c \propto M^{0.5}$. The halo mass required to host on average
one satellite LRG above a luminosity threshold is found to be about 10 times 
higher than that required to host a central LRG above the same threshold. 
We find that in massive halos the distribution of $L_*$ galaxies roughly 
follows that of the dark matter and their mean occupation number scales
with halo mass as $M^{1.5}$. The HOD modeling results also allows for 
an intuitive understanding of the scale-dependent luminosity dependence 
of the cross-correlation between LRGs and $L_*$ galaxies. Constraints
on the LRG HOD provide tests to models of formation and evolution 
of massive galaxies, and they are also useful for cosmological parameter 
investigations. In one of the appendices, we provide LRG HOD parameters with
dependence on cosmology inferred from modeling the two-point auto-correlation 
functions of LRGs.
\end{abstract}

\section{Introduction}

The clustering of galaxies depends on their properties, such as 
morphology 
(e.g.,
\citealt{hubble36,zwicky68,davis76,dressler80,postman84,guzzo97,willmer98,
zehavi02,goto03}), 
luminosity
(e.g., \citealt{davis88,hamilton88,white88,park94,loveday95,guzzo97,
benoist96,norberg01,zehavi02,Zehavi05b,Coil06,Coil08}), 
color 
(e.g., \citealt{willmer98,brown00,zehavi02,Zehavi05b,Coil08}), 
and spectral type 
(e.g., \citealt{norberg02,budavari03,madgwick03}).
Galaxy clustering thus provides important clues to
the physics of galaxy formation. Often found to reside in galaxy groups and 
clusters, luminous red galaxies (LRGs) constitute the bright end of the galaxy 
luminosity function. Clustering of LRGs encodes information about their 
environments, which are typically the central regions of groups and clusters.  
The LRG redshift sample \citep{Eisenstein01} of the Sloan Digital Sky Survey 
(SDSS; \citealt{York00}) provides an enormous data set with which to measure 
the clustering of LRGs. In this paper, we perform theoretical modeling of 
auto-clustering of LRGs and cross-clustering of LRGs with other types of 
galaxies to understand the origin of their clustering properties, to learn 
how they are distributed among massive dark matter halos, and to aid the 
study of formation and evolution of massive galaxies.

The theoretical understanding of galaxy clustering has been greatly enhanced
through the framework of the halo occupation distribution 
(HOD, see, e.g., \citealt{Jing98a,Ma00,Peacock00,Seljak00,Scoccimarro01,
Berlind02}) and the closely related approach of the conditional luminosity 
function (CLF, \citealt{Yang03}). The HOD formalism describes the bias 
relation between galaxies and matter at the level of individual virialized 
dark matter halos, whose distribution and properties can be readily predicted
by numerical simulations or analytic models given a cosmological model. 
The key ingredients of this formalism are the probability distribution 
$P(N|M)$ that a halo of mass $M$ contains $N$ galaxies of a given type
and spatial and velocity distributions of galaxies within halos. In the 
CLF approach, the dependence of $P(N|M)$ on galaxy luminosity is implicitly
derived by inferring the conditional luminosity distribution of galaxies 
as a function of halo mass $M$. Given the 
HOD/CLF and the cosmology, any statistics of galaxy clustering on any 
scales can be predicted, therefore the HOD/CLF method provides a complete 
description of the bias relation between galaxies and dark matter. 
HOD/CLF modeling has been applied to interpret galaxy clustering
measurements in several galaxy surveys (e.g., \citealt{Jing98b,Jing02,
Bullock02,Moustakas02,Bosch03,Magliocchetti03,Yan03,Zheng04,Yang05,Zehavi05b,
Lee06,Hamana06,Cooray06,Conroy06,White07,Zheng07,Blake08,Wake08}). HOD/CLF 
analysis recasts galaxy clustering measurements into a form that is more 
physically informative and conducive for testing galaxy formation theories 
(see, e.g., \citealt{Berlind02,Berlind03,Bosch03,Zheng05}).
\citet{Zehavi05b} present clustering measurements and HOD modeling
for galaxies in the SDSS main galaxy
spectroscopic sample \citep{Strauss02}. The 
HOD modeling results 
of the luminosity and color dependence of galaxy clustering 
are found to be in good qualitative 
agreement with predictions from galaxy formation models (\citealt{Zheng05}).

In this paper, we apply HOD modeling to interpret clustering in the SDSS
LRG spectroscopic sample, which uses color and magnitude cuts to effectively
select LRGs out to redshift $z\sim 0.45$ \citep{Eisenstein01}. The large
volume probed by the LRG sample has led to the detection of the baryon 
acoustic peak in the two-point correlation function 
(\citealt{Eisenstein05b}). \citet{Zehavi05a} report the measurements of
two-point correlation functions of 35,000 LRGs on scales of 0.3--40$\hMpc$. 
They find that LRGs are highly clustered (correlation length $\sim$10$\hMpc$) 
and that more luminous LRGs are more clustered. Clear deviations from a 
power law are seen in the correlation functions, with a dip at $\sim$2$\hMpc$. 
\citet{Eisenstein05a} measure the two-point cross-correlation between 
32,000 spectroscopic LRGs and 16 million galaxies in the SDSS imaging 
sample. Since these reference galaxies have luminosities around 
$L_*$, the characteristic luminosity of the \citet{Schechter76} luminosity
function, they are denoted as $L_*$ galaxies. \citet{Eisenstein05a} find a 
strong luminosity dependence of the LRG--$L_*$ cross-clustering amplitude. 
The form of the luminosity dependence 
is itself dependent on scale, 
with more variation in the 
clustering amplitude on small scales. Understanding all these auto- and 
cross-clustering features is one of the goals of our HOD modeling. Since 
LRGs trace massive halos, the cross-correlation between LRGs and $L_*$
galaxies also allows us to study the HOD of $L_*$ galaxies in these massive
halos.

The structure of this paper is as follows. In \S~2, we briefly describe the 
SDSS LRG and $L_*$ galaxy samples we use and our modeling method.
In \S~3, we describe our HOD parameterization for LRG samples and $L_*$ 
samples. In \S~4, we perform HOD modeling for LRG two-point auto-correlation 
functions and the two-point cross-correlation functions between LRGs and 
$L_*$ galaxies. We show how LRGs occupy dark matter halos and how $L_*$ 
galaxies occupy massive halos. Based on the modeling results, we interpret 
the luminosity dependence of the cross-clustering. Finally, we summarize 
and discuss our results in \S~5. We also include three appendices in the 
paper. In Appendix~\ref{sec:appendixA}, we investigate which HOD parameter 
plays the major role in the departures of the galaxy two-point correlation 
function from a pure power law. In Appendix~\ref{sec:appendixB}, we present 
HOD modeling results for two luminosity threshold LRG samples for different 
cosmological models. In Appendix~\ref{sec:appendixC}, we have a brief 
discussion on the mass function of the most massive halos and the fluctuation 
of the number of massive halos in the volume probed by the SDSS LRG samples. 

\section{Samples and Method}

The LRG two-point auto-correlation functions in the SDSS have been measured 
by \citet{Zehavi05a} with spectroscopic samples. The luminosity cuts of the 
two LRG samples at $z\sim0.3$ we model in this paper are $-23.2<M_g<-21.2$ 
and $-23.2<M_g<-21.8$, where $M_g$ is the restframe $g$-band absolute 
magnitude at $z=0.3$ computed from the observed $r$-band magnitude with
$k$ and passive evolution corrections \citep{Eisenstein05a}. Since galaxies 
with $M_g<-23.2$ are extremely rare, 
these can be regarded essentially as luminosity-threshold samples. For 
brevity, we call them the $M_g<-21.2$ sample and the $M_g<-21.8$ sample, or 
simply the faint sample and the bright sample. The comoving number densities
of the two samples are $9.73\times 10^{-5}h^3{\rm Mpc^{-3}}$ and 
$2.40\times 10^{-5}h^3{\rm Mpc^{-3}}$, respectively. Because of the fiber 
collision effect, the smallest scale can be probed by the spectroscopic LRG 
samples is about 0.4$\hMpc$. \citet{Masjedi06} extend the measurements of 
the two-point correlation function for the $M_g<-21.2$ sample down to a scale
of $\sim$0.015$\hMpc$ by cross-correlating the spectroscopic sample with the 
imaging sample to avoid the fiber collision effect. They find that the 
real-space two-point correlation function of the $M_g<-21.2$ sample roughly
follow an $r^{-2}$ profile from $\sim$100$\hMpc$ down to $\sim$0.01$\hMpc$.
In this paper, we focus on the measurements from the spectroscopic samples
in \citet{Zehavi05a} and limit ourselves to modeling the clustering on scales
above 0.3$\hMpc$.

The two-point cross-correlation functions between spectroscopic LRG samples 
and the imaging $L_*$ galaxy sample in the SDSS at $z\sim 0.3$ are measured by 
\citet{Eisenstein05a}. The $L_*$ galaxy sample we model is the one defined 
by the luminosity range $M^*-0.5$ to $M^*+1.0$, which is called the $M^*+1.0$ 
sample in \citet{Eisenstein05a}. For the LRG sample in the cross-clustering, 
we adopt the luminosity-bin sample defined by $-21.7<M_g<-21.2$.

We essentially follow the method presented in \citet{Zehavi05b} and adopt
the improvements of \citet{Tinker05} for theoretical modeling of the 
two-point auto- and cross-correlation functions in the HOD framework. In 
this paper, more general HOD parameterizations are used, as described 
in the following sections. While the two-point auto-correlation functions 
in \citet{Zehavi05a} are projected along the redshift direction, similar 
to those in \citet{Zehavi05b}, the LRG--$L_*$ two-point cross-correlation 
functions $\Delta$ measured in \citet{Eisenstein05a} are volume-averaged 
real-space cross-correlation functions $\xi_\times$. The average is weighted 
by a spherical window function,
\begin{equation}
\label{eqn:Delta}
\Delta(a)=\frac{1}{V} \int_0^{\infty} dr 4\pi r^2  \xi_\times(r)W(r;a),
\end{equation}
where the window function is of the form
\begin{equation}
\label{eqn:window}
W(r;a) = \frac{r^2}{a^2} \exp\left(-\frac{r^2}{2a^2}\right).
\end{equation}
The effective volume $V$ for the window function is $3(2\pi a^2)^{3/2}$.
In our modeling, we perform the same volume average to compute the predicted
cross-correlation function. 

When performing fits to the two-point auto- and cross-correlation functions,
we calculate values of $\chi^2$ using the full error covariance matrices, 
inferred through the jackknife method (see \citealt{Eisenstein05a} and 
\citealt{Zehavi05b} for details).

Throughout the paper, we adopt the spatially flat ``concordance'' cosmological
model with the matter density parameter $\Omega_m=0.3$ and baryon density 
parameter $\Omega_b=0.047$. We assume adiabatic Gaussian primordial 
density fluctuations with a power-law index of the spectrum $n_s=1$.
The r.m.s. matter density fluctuation in spheres of radius $8\hMpc$ linearly 
extrapolated to $z=0$ is assumed to be $\sigma_8=0.8$. The Hubble constant 
we use is $h=0.7$ in unit of 100 km~s$^{-1}$~Mpc$^{-1}$. Appendix B presents 
the dependence of the derived HOD parameters on $\sigma_8$ and $\Omega_m$ for
a five-parameter HOD model. The mean redshift of the above LRG and $L_*$ 
samples is around 0.3 and our model calculations take this into account, 
i.e., all halo properties are calculated at $z=0.3$ for the adopted cosmology. 
In general, the comoving unit is adopted for distance, but to be compatible 
with \citet{Eisenstein05a}, we also use proper units with explicit mention 
when discussing the cross-clustering. 
Finally, we assume $h=1$ when  quoting magnitudes throughout the paper.
 
\section{HOD Parameterization}
\label{sec:hod}

\subsection{The Mean Occupation Function of LRGs}
\label{sec:meanlrg}

The two LRG samples for which the two-point auto-correlation functions are 
modeled are nearly luminosity-threshold samples. The LRG samples defined 
in the cross-correlation measurements in \citet{Eisenstein05a} are 
luminosity-bin samples. We concentrate on the cross-correlation between 
$L_*$ galaxies and $-21.7<M_g<-21.2$ LRGs, together with the two-point 
auto-correlation functions of the two luminosity-threshold samples
with $M_g<-21.2$ and $M_g<-21.8$. The three LRG samples with
different luminosity cuts provide leverage to constrain the relation 
between galaxy luminosity and halo mass. We therefore first parameterize 
the LRG HOD in a way that includes the luminosity information.

For HOD parameterization, it has been found to be useful to separate
contributions from central and satellite galaxies (\citealt{Kravtsov04,
Zheng05}).  Central galaxy luminosity is correlated with the host halo mass.
Based on predictions of galaxy formation models (\citealt{Zheng05}), 
we assume that at a fixed halo mass the central galaxy luminosity follows 
a log-normal distribution and the mean luminosity of this distribution has 
a power-law form $L_c=L_s(M/M_s)^p$, with $p$ independent of halo mass in 
the range that LRGs probe. In terms of absolute magnitudes,
\begin{equation}
\label{eqn:cenmag}
M_{gc}=M_{gs}-2.5p\log(M/M_s),
\end{equation}
where $M_{gs}$ is the mean luminosity in halos with a pivot mass $M_s$ and
is simply set to be $-19.8$. We note that in this paper $M$ with a subscript 
``{\it g}'' stands for the $g$-band absolute magnitude, while that without the 
subscript ``{\it g}'' is used for halo mass. With a constant standard deviation
$\sigma_{M_g}$ in magnitude, the luminosity distribution of central galaxies 
in halos of mass $M$ is then
\begin{equation}
\label{eqn:cenclf}
\frac{d\Ncen}{dM_g} = \frac{1}{\sqrt{2\pi}\sigma_{M_g}}
                     \exp\left[-\frac{(M_g-M_{gc})^2}{2\sigma_{M_g}^2}\right].
\end{equation}
LRG samples are not defined purely through luminosity, since there is a color 
cut in the LRG selection. Therefore, for an LRG sample with 
$M_g<M_{g,{\rm thres}}$, the mean occupation function of central galaxies is 
calculated as 
\begin{equation}
\label{eqn:Ncen}
\Ncen = \int_{-\infty}^{M_{g,{\rm thres}}} dM_g \frac{d\Ncen}{dM_g} f(M_g),
\end{equation}
where $f(M_g)$ is the LRG selection function at a given luminosity. Based 
on the LRG luminosity function and the total galaxy luminosity function,
we approximate the selection function as $f(M_g)=-2(M_g+20.45)/3$ in the
range $-21.95<M_g<-20.45$, 0 for $M_g>-20.45$, and 1 for $M_g<-21.95$.
The HOD of a luminosity-bin galaxy sample is just the difference of those of 
two luminosity-threshold samples. 

The theoretically predicted mean occupation function of satellite galaxies 
for a luminosity-threshold
sample is close to a power law (\citealt{Kravtsov04,Zheng05}). Here, we adopt 
a more flexible parameterization, which can give us a better idea on
the constraining power of the two-point clustering on the HOD. For each 
luminosity-threshold LRG sample, 
we parameterize the HOD of satellite LRGs through the mean occupation numbers 
at five mass scales, and the the mean occupation function is assumed to be a 
cubic spline curve passing through the five points (see \citealt{Zheng07a}). 
Linear extrapolations are used outside of the ranges of the five points. In
addition, 
we impose the central galaxy cutoff profile $\Ncen$ to $\Nsat$. The five 
mass scales are chosen to be $\log (M/\hMsun) = 14.10+0.35i$ 
($i=0,1,2,3,4$). 
The mean occupation function of satellites in 
a luminosity-bin sample is obtained by interpolating the values of 
$\log \Nsat$ for the two luminosity-threshold samples (linearly in $\log L$).
The satellite occupation number around the mean is assumed to follow
the Poisson distribution (\citealt{Kravtsov04,Zheng05}).

The parameterization of the halo occupation of LRGs we present here is
a hybrid of the CLF framework (\citealt{Yang03}) and the HOD framework
(\citealt{Berlind02}) --- the distribution of central LRGs is put in a CLF
form, while that of the satellite LRGs is in the usual HOD form. This
combination allows us to infer information on how luminosity of central
galaxies changes with halo mass and at the same time to keep a reasonable
number of free parameters.

With the above parameterized HOD, we can easily form the mean occupation
functions for the two luminosity-threshold LRG samples and the luminosity-bin
sample given their luminosity cuts. The thirteen free parameters are the pivot
mass scale $M_s$ and the slope $p$ in the luminosity-mass relation, the
scatter $\sigma_{M_g}$ of the luminosity distribution of central LRGs, 
and the total of ten spline points of $\Nsat$ for the two luminosity-threshold 
samples. In Appendix~\ref{sec:appendixB}, we provide HOD modeling results 
for the two luminosity threshold samples and their cosmology dependence 
based on a simpler five-parameter HOD model.

\subsection{The Mean Occupation Function of $L_*$ Galaxies}

The $L_*$ sample in \citet{Eisenstein05a} is at a mean redshift $z\sim0.3$.
To parameterize the HOD of this $L_*$ sample, we make use of the modeling
result in \citet{Zehavi05b} at $z\sim 0$. The $L_*$ galaxy sample is a
luminosity-bin sample $M^*-0.5<M_r<M^*+1.0$, which can be regarded as the 
difference between two luminosity-threshold samples, $M_r<M^*+1.0$ and
$M_r<M^*-0.5$. As shown in \citet{Kravtsov04}, at low redshifts, for threshold
samples of a fixed (comoving) galaxy number density, the shape of the mean 
occupation function as a function of redshift approximately remains unchanged.
We take the advantage of this presumed property and construct the mean occupation 
function for $L_*$ galaxies at $z\sim0.3$ based on the HODs of $z\sim 0$ 
galaxies. We first find out the number densities of the two 
luminosity-threshold samples with $M_r<M^*+1.0$ and $M_r<M^*-0.5$ at
$z\sim 0.3$. To do this, we adopt the evolution model of the 
luminosity function in \citet{Blanton03} set to $z=0.3$ and obtain the two
number densities as $n_1=1.1\times 10^{-2}h^3{\rm Mpc^{-3}}$ and 
$n_2=1.3\times 10^{-3} h^3{\rm Mpc^{-3}}$, respectively.
We find the luminosity thresholds $L_1$ and $L_2$ for $z\sim 0$ galaxies 
that match the two number densities,
\begin{equation}
n(L>L_1; z\sim 0)=n_1
\end{equation}
and
\begin{equation}
n(L>L_2; z\sim 0)=n_2.
\end{equation}
We then  use the results presented in 
\citet{Zehavi05b} to infer the HODs at $z\sim 0$ for luminosity-threshold 
samples with $L_1$ and $L_2$ as the thresholds. Finally, we take the 
difference of these two HODs to infer the mean occupation function for 
galaxies in the $L_1<L<L_2$ luminosity bin,
\begin{equation}
\langle N(L_1<L<L_2; M)\rangle = 
\langle N(L>L_1; M)\rangle - \langle N(L>L_2; M)\rangle.
\end{equation}
We take the {\it shape} of this mean occupation function as that for 
the $L_*$ galaxies at 
$z\sim0.3$. The form of this  mean occupation function is a sum of a square 
window for central galaxies and a 
power law with an index $1.10$ for satellites. With the shape fixed,
we allow the mean occupation function to shift in mass scales. Furthermore, 
we add freedom to the high mass slope, as described below, because we 
have better data to constrain it.

As shown later, the small-scale two-point cross-correlation between $L_*$ 
galaxies and LRGs comes from satellite $L_*$ galaxies paired with LRGs.
As our goal here is to investigate the $L_*$ occupation in high mass halos,
it is necessary to introduce additional degrees of freedom in the mean 
occupation function of the satellite $L_*$ galaxies in massive halos.
We only adopt the $z\sim0.3$ mean occupation function for $L_*$ satellites,
as constructed above, up to $10^{13}\hMsun$, and introduce another power 
law to represent the mean occupation function of (satellite) $L_*$ galaxies 
in halos more massive than $10^{13}\hMsun$.  We assume this occupation 
distribution follows the Poisson distribution with the mean in the form of
$N_t(M/M_t)^{\alpha_h}$. Here $M_t$ is a pivot mass scale fixed
to be $2.5\times10^{14}\hMsun$, $N_t$ is the mean occupation number of
$L_*$ galaxies in halos of this pivot mass, and $\alpha_h$ denotes the
slope at the high mass end. The scale of the pivot mass is chosen to
minimize the correlation between $N_t$ and $\alpha_h$. In order to have
more flexibility, we do not impose continuity at $10^{13}\hMsun$. 

At first glance, our construction of the $L_*$ HOD seems complex,
involving theoretical priors and introducing an artificial break in the 
satellite occupation function. A more general HOD model for $L_*$ galaxies 
could be obtained by parameterizing the HODs of two luminosity threshold 
samples and taking their difference (e.g., \citealt{Tinker07}). However, we 
only have a small number of data points from the LRG-$L_*$ cross-correlation 
functions to constrain the $L_*$ HOD. The constraints on the low mass part of 
the $L_*$ HOD (especially the mean occupation function for central galaxies) 
come from the the number density of the $L_*$ sample and the large scale bias 
of the cross-correlation. The shape of the low mass part of a general $L_*$ 
HOD can not be well constrained by either the number density or the large 
scale bias, given that the halo bias factor in this mass range only increases 
slowly. The constraints to the high mass part of the $L_*$ HOD come from the
small-scale cross-correlation with the LRGs. 
Our procedure of the $L_*$ HOD parameterization reduces the degrees of freedom
in the low mass part of the HOD and allows some flexibility in the high
mass part of the HOD which is our focus. Although our parameterization
is a restricted one, it is suitable for our purpose of understanding 
clustering properties in $L_*$-LRG cross-correlation. 
To summarize, we allow the low mass end of the mean occupation function 
(a square window plus 
the low-mass power law) to have an overall horizontal shift with the shape 
fixed. The main role of shifting the low mass part of the $L_*$ HOD is 
to match the number density of the $L_*$ sample and the large scale bias 
of the cross-correlation, while the high mass part of the occupation function 
of $L_*$ galaxies that we are interested in determines the small-scale 
cross-correlation with the LRGs.

The distribution profile of $L_*$ satellite galaxies inside halos is assumed
to follow the Navarro-Frenk-White (NFW) profile (\citealt{Navarro95,
Navarro96,Navarro97}) with a concentration parameter 
$c=c_0(M/M_{\rm nl})^\beta$, where $M_{\rm nl}\sim 1.5\times 10^{12}\hMsun$ 
is the nonlinear mass at $z\sim 0.3$ and $\beta$ is fixed to -0.13 
(\citealt{Bullock01}). As a whole, there are four free parameters in the 
$L_*$ HOD, the overall horizontal shift of the low mass part, the high mass 
amplitude $N_t$ and slope $\alpha_h$ of the mean occupation function, and 
the halo concentration parameter $c_0$ extrapolated to the nonlinear mass
scale.
To compute the one-halo term of LRG--$L_*$ cross-correlation, we make the
simplifying assumption that the occupation number of LRGs and that of $L_*$ 
galaxies inside the same halo are uncorrelated. That is, 
$\langle N({\rm LRG};M) N(L_*;M)\rangle = 
 \langle N({\rm LRG};M) \rangle
 \langle N(L_*;M)\rangle$.
We discuss the effect of this assumption on our modeling results in \S~4.

\begin{figure*}
\epsscale{1.0}
\plotone{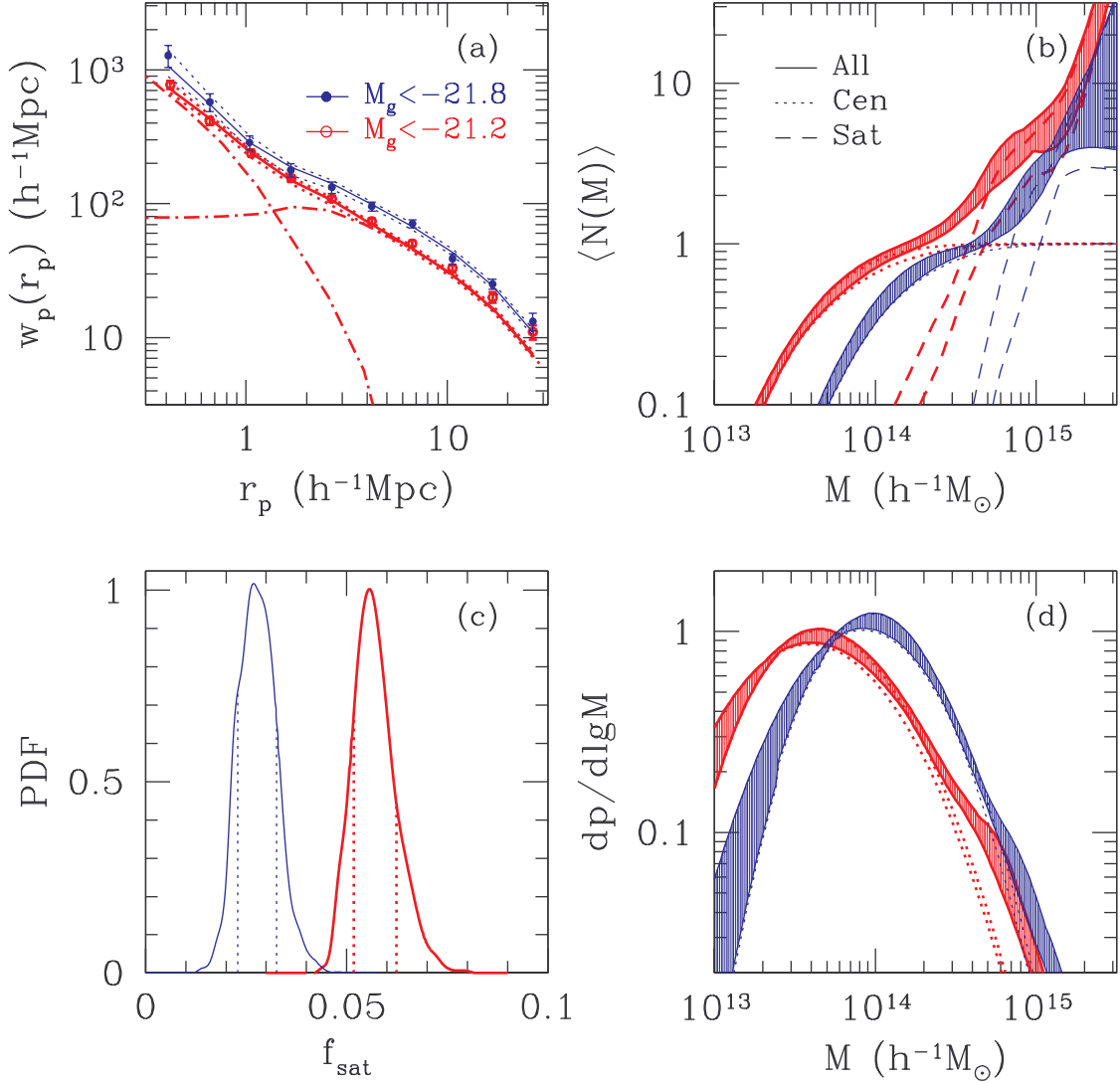}
\epsscale{1}
\caption[]{
\label{fig:wp_LRG}
Projected two-point auto-correlation functions and best-fit HODs for the 
two luminosity-threshold LRG samples. 
Panel ({\it a}): 
      The measured two-point correlation functions (data points and error 
      bars) and the HOD model fits (solid curves). The two dashed curves for 
      each sample show the envelope of predictions from models with 
      $\Delta\chi^2<4$. The predicted one-halo and two-halo terms (dot-dashed 
      curves) are also shown for the sample with the lower luminosity 
      threshold.
Panel ({\it b}): 
      The mean occupation functions (solid curves) of LRGs from the best fits, 
      with contributions from central (dotted) and satellite (dashed) LRGs. 
      For each sample, the two sets of curves are the envelope from models 
      with $\Delta\chi^2<4$ 
and those for all galaxies are shaded.
Panel ({\it c}): 
      The marginalized distribution of the satellite fraction in each sample 
      with the central 68\% distribution marked by the two dashed lines.  
Panel ({\it d}): 
      The probability distribution of halo masses for the LRGs in each sample 
      (solid lines), obtained from the occupation function shown in panel 
      ({\it b}) weighted by the differential halo mass function. The dotted 
      lines show the halo mass probability distribution from just the central 
      galaxies.
}
\end{figure*}

\section{Modeling Results} 

We model simultaneously  the two-point auto-correlation functions of the two 
luminosity-threshold LRG samples ($M_g<-21.2$ and $M_g<-21.8$)  and the 
two-point cross-correlation function between the $L_*$ and the luminosity-bin 
($-21.7<M_g<-21.2$) LRG sample. Altogether, there are seventeen free 
parameters, four for the $L_*$ HOD and thirteen for the LRG HOD. 

In addition 
to the sum of $\chi^2$s from the two auto-correlation functions and the
cross-correlation function, we also add the number densities of the two 
threshold LRG samples and the $L_*$ galaxies into the overall $\chi^2$.
That is, 
\begin{eqnarray}
\chi^2 & = & \mathbf{(w_1-w_1^*)^T C_1^{-1} (w_1-w_1^*)} \nonumber \\
       & + & \mathbf{(w_2-w_2^*)^T C_2^{-1} (w_2-w_2^*)} \nonumber \\
       & + & \mathbf{(\Delta-\Delta^*)^T C_\times^{-1} (\Delta-\Delta^*)} 
       +\sum_{i=1}^3\frac{(n_i-n_i^*)^2}{\sigma_{n_i}^2},
\end{eqnarray}
where $\mathbf{w_1}$, $\mathbf{w_2}$, and $\mathbf{\Delta}$ are the vectors of
auto-correlation functions of the two LRG samples and the LRG--$L_*$ 
cross-correlation function, and $n_i$ ($i=1,2,3$) are the three number 
densities. The observed values are denoted with a superscript $*$. 
The full covariance matrix is used for each correlation function and 
10\% fractional errors are assumed for each of the galaxy number densities.

The three covariance matrices, $\mathbf{C_1}$, $\mathbf{C_2}$, and 
$\mathbf{C_\times}$ are estimated using jackknife resampling. For each of 
the two LRG samples, 104 jackknife subsamples are used \citep{Zehavi05a} for
estimating the covariance among 10 data points. For the LRG-$L_*$ 
cross-correlation, the covariance among 6 data points is estimated with
50 jackknife subsamples \citep{Eisenstein05a}. 
\citet{Zehavi05b}  performed extensive tests with mock catalogs to access
the reliability of the jackknife error estimates in projected correlation
functions over a similar range of separations. They used 100 mock catalogs 
with the same geometry and angular completeness as the SDSS sample and 
similar clustering properties, created using the PTHalos method of 
\citet{Scoccimarro02}. Their tests showed that
the jackknife method is a robust way to estimate the error covariance matrix,
especially for the relatively large volumes probed here. This holds
as long as the number of jackknife realizations, $n$, is significantly 
larger than the dimension of the data vector, $p$. 
\citet{Hartlap07} discusses related issues, pointing out a potential bias
in general model fitting which depends on the ratio of $p$ to $n$. As a
crude test of our error uncertainties, we incorporate their proposed method
to remove this bias by multiplying the inverse of the covariance matrix by 
a factor of $(n-p-2)/(n-1)$.  In our case, this factor is about 0.9. We
perform this for the $M_g<21.2$ LRG sample, resulting in a $\sim10\%$
increase in the uncertainties of the HOD parameters, but no noticeable
change in the best-fit values.  We note 
that the results presented later in this paper do not include such a 
correction.

We assume the different clustering measurements to be independent, ignoring
possible correlations between statistical errors across correlation 
functions and number densities of different samples. Strictly speaking, 
such statistical 
correlations are not zero. For example, the $M_g<-21.2$ LRG sample includes
LRGs with $M_g<-21.8$, therefore the auto-correlation functions of the
$M_g<-21.2$ and $M_g<-21.8$ samples are partially correlated. These 
correlations could be estimated with the jackknife technique, but they would
be noisy. Neglecting such correlations might make the constraints on the 
HOD parameters somewhat tighter than they should be, which is a caveat for 
interpreting our results.

We adopt a Markov Chain Monte Carlo technique (MCMC;
e.g., \citealt{Gilks96}) to explore the HOD parameter space. At each point 
of the chain, we take a random walk in the parameter space to generate a 
new set of HOD parameters. The step-size of the random walk for each parameter
is drawn from a Gaussian distribution. The probability to accept the new set 
of HOD parameters depends on the difference between $\chi^2_{\rm old}$ and 
$\chi^2_{\rm new}$ (the values of $\chi^2$ for the new and old models): 1 
for $\chi^2_{\rm new}\leq\chi^2_{\rm old}$ and $\exp[-(\chi^2_{\rm new}-
\chi^2_{\rm old})]$ for $\chi^2_{\rm new}>\chi^2_{\rm old}$. 
Flat priors in logarithmic space are adopted for the LRG satellite occupation 
numbers at the ten spline points and the pivot mass scale $M_s$ and for $N_t$ 
and mass scale shift of $L_*$ galaxies. Flat priors in linear space are used
for other HOD parameters. 

The total number of data points to model is 29 (10+10 auto-correlation 
measurements for the two LRG samples, 6 LRG-$L_*$ cross-correlation 
measurements, and 3 number densities). The number densities are well 
reproduced by our model, with the median values 1.1-$\sigma$, 0.4-$\sigma$, 
and 0.1-$\sigma$ away from the observed ones for the two LRG and the $L_*$ 
samples. Our model has 17 free parameters
and therefore the number of degree of freedom is 12. We find that the 
best-fit model has $\chi^2=25$. The probability for a $\chi^2$ value higher 
than our best-fit  $\chi^2$ is 1.5\%. (Including the \citet{Hartlap07}
correction gives $\chi^2 \sim 22$ and the probability increases to $\sim$4\%). 
The relatively-large $\chi^2$ values might be partly caused by our neglecting 
the correlation between statistical errors across correlation functions and 
number densities of different samples. It also indicates that the accuracy 
in our analytical model of the two-point galaxy correlation function needs 
to be improved and that our HOD parameterization is not perfect. With the 
above caveats in mind, we present our modeling results below.

\begin{figure*}[t]
\epsscale{1.0} 
\plotone{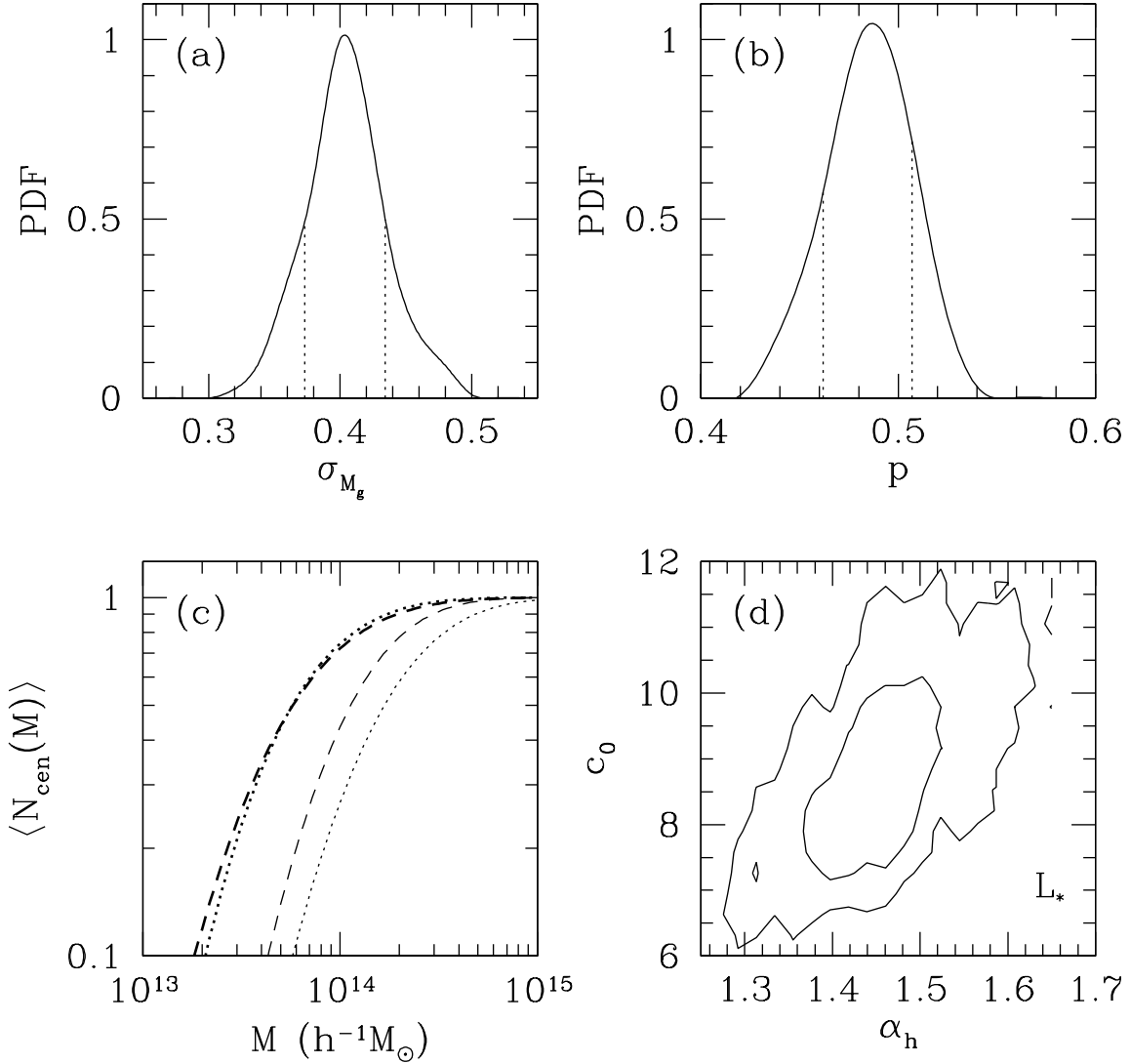}
\epsscale{1.0}
\caption[]{
\label{fig:crosspar}
Constraints on  HOD parameters from the LRG auto-correlation functions and
the LRG-$L_*$ cross-correlation function (see the text for details).
Panel ({\it a}): 
      The marginalized distribution of the parameter $\sigma_{M_g}$, which 
      is the width (in magnitude) of central galaxy luminosity distribution 
      at a fixed halo mass.  
Panel ({\it b}): 
      The marginalized distribution of the parameter $p$, which characterizes 
      the relation between mean central galaxy luminosity and halo mass 
      through $L_c\propto M^p$.
In panels ({\it a}) and ({\it b}), the two dashed vertical lines
indicate the central 68\% of the distribution.
Panel ({\it c}): 
      Illustration of the effect of $\sigma_{M_g}$ and $p$ on the mean 
      occupation functions of LRGs. The thick dotted and dashed curves 
      are for the $M_g<-21.2$ LRG sample, with $\sigma_{M_g}$ varied from 0.37 
      to 0.43 (i.e., a $\pm 1\sigma$ change). The thin dashed and dotted curves 
      are for the $M_g<-21.8$ LRG sample, with $p$ varied from 0.46 to 0.51 
      (a $\pm 1\sigma$ change in $p$).
Panel ({\it d}):  
      The marginalized joint distribution of the concentration parameter 
      $c_0$ (normalized to that at the $z\sim 0.3$ nonlinear mass scale) 
      and the high mass slope $\alpha_h$ of the mean occupation function for 
      the $L_*$ galaxies. The contours show the 68\% and 95\% confidence 
      levels for two parameters.
}
\end{figure*}

\subsection{Constraints on the HOD of LRGs}

\subsubsection{HOD for the Luminosity-threshold LRG Samples}

Figure~\ref{fig:wp_LRG} shows the fitting results for the two 
luminosity-threshold LRG samples.  Figure~\ref{fig:wp_LRG}{\it a} shows 
the best-fit projected two-point correlation functions together with 
the measurements.  For the $M_g<-21.2$ sample, the one-halo and two-halo 
components (dot-dashed curves) of the fit are also shown.  In 
Figure~\ref{fig:wp_LRG}{\it b}, the mean occupation function for each 
sample is plotted, separating into contributions from central (dotted) and 
satellite (dashed) galaxies.  
The range of mean occupation functions with $\Delta\chi^2<4$ is denoted
by the shaded region ($\Delta\chi^2<4$ is chosen so that the envelopes are
sampled by a large number of MCMC points and the range delineated by the
envelopes can be clearly seen in the plot).
As in \citet{Zehavi05b}, the amplitude of the 
high halo mass end of the mean occupation function is poorly constrained by 
two-point correlation functions. Since the abundance of halos drops 
exponentially at the high mass end, the two-halo pairs are mostly contributed 
by lower mass halos. Although the number of one-halo pairs per halo rises 
roughly as $\NavgM^2$, the exponential drop of the number of high mass halos 
also makes most one-halo pairs come from lower mass halos. Overall, the 
two-point correlation function is dominated by signals from halos of lower 
mass where $\NavgM$$\sim$ a few, leading to the poor constraints on the HOD 
at the very high mass end. The constraint on the high mass end 
{\it slope} of $\Nsat$ for the faint LRG sample appears to be relatively
strong, and we discuss its possible implications in 
\S~\ref{sec:discuss}.

The results show that the host halos of LRGs are massive, above 
$10^{13}\hMsun$, corresponding to large galaxy groups and clusters,
which is consistent with studies of the environment of luminous galaxies
(e.g., \citealt{Loh2003}). We find that as the threshold luminosity increases 
by a factor of $\sim$1.7, the mass scale of host halos shifts by a factor of 
$\sim 2.3$, which implies a correlation between the luminosity $L_c$ of the 
central LRG and the mass $M$ of the host halo in the form of $L_c \propto 
M^{0.66}$.

LRGs are often thought to be the central elliptical galaxies of galaxy 
groups of clusters. Our results clearly show that a fraction of the LRGs 
have to be satellite galaxies. The main constraint for this comes from the 
high amplitudes of the correlation function at small scales. Without 
satellite LRGs, the projected correlation function would  flatten out
toward small scales, similar in shape to the two-halo term shown in 
Figure~\ref{fig:wp_LRG}{\it a}. To explain the rising small-scale 
clustering, the one-halo term has to be introduced, hence we require the
existence of satellite LRGs. However, the fits imply that only a small 
fraction of LRGs are satellites: 5.2--6.2\% and 2.3--3.2\% 
in the $M_g<-21.2$ and $M_g<-21.8$ samples, respectively, as shown in 
Figure~\ref{fig:wp_LRG}{\it c}. The steeper rise of the small scale 
clustering makes the two-point correlation function of the 
brighter LRG sample deviate from a power law more prominently. In 
Appendix~\ref{sec:appendixA}, we investigate the key ingredients that 
lead to the departure from a power law and provide insight as to 
why the departure 
becomes more clear for samples that are more luminous (as shown here) and 
for samples at higher redshifts (e.g., \citealt{Ouchi05}).

The mean occupation function tells us the mean number of LRGs as a function
of the halo mass. The result can be viewed differently by asking what the
probability distribution of masses of halos hosting such LRGs would be.
The probability is simply the product of the mean occupation function and
the differential halo mass function. We show such probability distributions 
from the MCMC run in Figure~\ref{fig:wp_LRG}{\it d}.
The distribution of $M_g<-21.2$ LRGs peaks at $\sim 4.5\times 10^{13}\hMsun$, 
and for $M_g<-21.8$ LRGs the peak shifts to higher mass, $\sim 10^{14}\hMsun$.
Both distributions span a large range in halo mass. The full width at half 
maximum (FWHM) for either distribution is about $\Delta\log M=0.8$.  Since 
only a small fraction of LRGs are satellites, the probability distribution 
is almost determined by the central LRGs. Only in halos more massive than 
a few times $10^{14}\hMsun$ does the number of satellites in a halo become 
significant (i.e., greater than one on average), and around this mass scale 
a low amplitude shoulder in the distribution emerges. In FWHM sense, we 
find that $M_g<-21.2$ and $M_g<-21.8$ LRGs reside in 
halos of mass 2--13$\times 10^{13}\hMsun$ and 4--25$\times 10^{13}\hMsun$, 
respectively.

From the mean occupation functions in 
Figure~\ref{fig:wp_LRG}{\it b}, 
one also notices that, compared with the samples of lower luminosities
in \citet{Zehavi05b}, the low mass cutoff profiles for the LRG samples
are better constrained. The relatively tight constraints come from 
the steepening both in the halo mass function and in the mass dependence of
the halo bias factor 
toward high halo mass. The former steepening makes the galaxy number 
density more sensitive to the cutoff profile, while the latter one 
increases the sensitivity of the large-scale galaxy bias factor to the 
cutoff profile. With our HOD parameterization, the cutoff profile 
of the LRG mean occupation function encodes information on the distribution 
of central galaxy luminosity in halos of fixed mass and on how the mean 
central galaxy luminosity scales with halo mass. 
Figure~\ref{fig:crosspar}{\it a} shows the marginalized distribution of the 
width $\sigma_{M_g}$ (in magnitude) of the (log-normal) central galaxy 
luminosity distribution at a fixed halo mass. The clustering data require
a scatter of $\sim 0.16$ dex in the central galaxy luminosity in halos of
a given mass. The thick dotted and dashed curves in 
Figure~\ref{fig:crosspar}{\it c} shows how $\sigma_{M_g}$ affects the 
cutoff profile in $\Ncen$ of the $M_g<-21.2$ sample by varying 
$\sigma_{M_g}$ by $\pm 1\sigma$. We note that the constraints on the 
cutoff profile (and therefore $\sigma_{M_g}$) depend on cosmological
parameters, especially $\sigma_8$, in the sense of a larger $\sigma_{M_g}$ 
for a larger $\sigma_8$ (see Appendix~\ref{sec:appendixB} for more details). 

The mean central galaxy luminosity scales with halo mass, which is 
characterized by the parameter $p$, $L_c\propto M^p$.  LRG samples with 
different luminosity cuts allow us to constrain the parameter $p$. The 
marginalized distribution of $p$ is shown in Figure~\ref{fig:crosspar}{\it b}. 
The thin dotted and dashed curves in Figure~\ref{fig:crosspar}{\it c} show 
the effect $p$ on the mass scale shift of the $M_g<-21.8$ sample relative 
to the $M_g<-21.2$ sample. The value of $p$ is around 0.48, which seems to 
be inconsistent with the value 0.66 estimated from comparing the luminosity 
and halo mass scale of the two threshold LRG samples. The reason is simple --- 
in our parameterization, the parameters $p$ and $\sigma_{M_g}$ constrained 
here correspond to all central galaxies, not only central LRG galaxies that
have a color selection criterion imposed. 

Our HOD parameterization here is rather flexible in the satellite HOD.
Again, in Appendix~\ref{sec:appendixB}, we present modeling results with 
a five-parameter 
HOD model and show their dependence on cosmology. In \S~\ref{sec:discuss}, 
we compare our HOD modeling results with those from other works and discuss 
a few issues related to the modeling.

We note that, on large scales, the bestfit $w_p$ curve in 
Figure~\ref{fig:wp_LRG}{\it a} appears to be lower than that from the 
five-parameter model (Fig.~\ref{fig:5par_sig8}{\it a}) for the $M_g<-21.2$ 
sample. The low mass cutoff profiles of $\Ncen$ for the bright and the faint 
LRG samples are correlated in the parameterization adopted here, which means 
that the HOD for central galaxies is more restrictive than that in the 
five-parameter model. The LRG-$L_*$ cross-correlation also limits the range 
of LRG HODs. The bestfit solution is a compromise in matching the large-scale
amplitudes of both the LRG auto-correlation functions and the LRG-$L_*$ 
cross-correlation function. The result implies that our HOD parameterization 
is not perfect and that there is room to improve it (e.g., by allowing the 
scatter in the central galaxy luminosity to vary with halo mass).

\subsubsection{Mass Scales of Host Halos of Central and Satellite LRGs}

Applying HOD modeling to galaxy samples with different luminosity thresholds,
\citet{Zehavi05b} find that there is a remarkable scaling relation between
the characteristic minimum mass $\Mmin$ of the host halos and the mass scale
$M_1$ of halos that on average host one satellite galaxy (above the luminosity
threshold) in addition to the central galaxy, $M_1\sim 23\Mmin$. With a
HOD parameterization close to what is used here, \citet{Zheng07} found the 
relation is more like $M_1\sim 18\Mmin$.  
Given the parameterization described in \S\ref{sec:meanlrg}, we
define $\Mmin$ to be the halo mass at which the expected number
of central galaxies above the luminosity threshold is 0.5, whether
or not the galaxy satisfies the LRG color criteria.
Theoretical studies of HODs of 
subhalos in dissipationless dark matter simulations (\citealt{Kravtsov04}) 
and those predicted by SPH and semi-analytic galaxy formation 
models (\citealt{Zheng05}) reveal a similar relation with the scaling factor 
around 20. As shown by \cite{Berlind03}, the large gap between $M_1$ and 
$\Mmin$ arises because in the low occupation regime, a more massive halo
tends to host a more massive central galaxy, rather than multiple smaller 
galaxies. Does the $M_1$-$\Mmin$ scaling relation extend to massive halos 
hosting LRGs? 

\begin{figure}
\epsscale{1.0}
\plotone{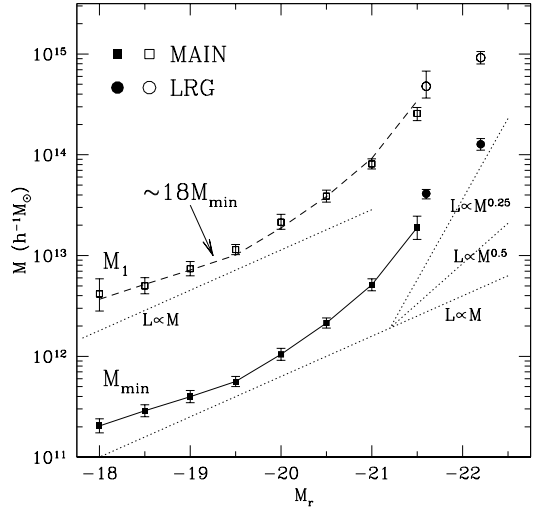}
\epsscale{1.0}
\caption[]{
\label{fig:ML}
Mass scales of the LRG HODs as a function of threshold luminosity.  Shown 
are the relation between the characteristic minimum mass $\Mmin$ 
at which 50\% of halos host central galaxies above the luminosity threshold
and the mass $M_1$ of halos that on average host one 
satellite galaxy, as a function of the threshold absolute magnitude. The square
points are taken from \citet{Zheng07} for the SDSS main galaxy sample 
(corrected to $\sigma_8=0.8$). Open and filled circles are the results 
for the two LRG samples (note that the $z=0.3$ $g$-band luminosity 
is converted to $z=0.1$ $r$-band by adopting an apparent color of $g-r=0.4$). 
Dotted lines indicate different relations between luminosity of central galaxy 
and mass of host halo.} 
\end{figure}

In Figure~\ref{fig:ML}, we plot $\Mmin$ and $M_1$ as a function of the 
threshold luminosity from \citet{Zheng07}, corrected to be consistent
with $\sigma_8=0.8$ adopted in this paper. The luminosity in \citet{Zheng07} 
is in $z=0.1$ $r$-band (see \citealt{Zehavi05b}). For comparison, we convert 
the $K$-corrected and passively evolved $z=0.3$ $g$-band threshold 
luminosities of the two LRG samples to the $z=0.1$ $r$-band ones by adopting 
an apparent color $g-r=0.4$, and we obtain $M_r<-21.6$ and $M_r<-22.2$, 
respectively.
We obtain the distribution of $\Mmin$ and $M_1$ for the two 
luminosity-threshold LRG samples modeled in this paper by solving 
$\Ncen=0.5$ and $\Nsat=1$ for each set of HOD parameters in the MCMC chain. 
The results 
are shown as the last two pairs of points. Roughly 
speaking, they seem to follow the previous trend. 
The mild discontinuity probably reflects an imperfect magnitude
conversion, which is not surprising as we are trying to account
for filter difference, $K$-correction, and stellar population evolution
from $z\sim 0.3$ to $z\sim 0.15$.
A more interesting difference is that the scaling 
factors between $M_1$ and $\Mmin$ become 10.5$_{-1.5}^{+5.1}$ ($M_g<-21.2$) 
and 7.2$_{-0.8}^{+0.9}$ ($M_g<-21.8$), respectively. 
For comparison, the mean $M_1$-to-$\Mmin$ ratio for the MAIN galaxy samples 
shown in Figure~\ref{fig:ML} is 18.3 with a mean 1$\sigma$ uncertainty of 
$\sim$3.6. 
This indicates that 
the scaling relation may 
break down in very massive halos. 
In fact, the scaling relation from the 
MAIN sample with the highest luminosity threshold ($M_r<-21.5$) in 
Figure~\ref{fig:ML} has already shown a trend of such a decrease, with an
$M_1$-to-$\Mmin$ ratio of 13.2$_{-2.4}^{+2.7}$.
As a whole, these results show that the 
$M_1$-$\Mmin$ scaling factor decreases for luminous galaxies that reside in
massive halos. We caution, however, that the definition of the $M_1/\Mmin$
ratio becomes more parameterization dependent at high galaxy luminosities
because the scatter between luminosity and halo mass is larger (i.e., the low 
mass cutoff of the HOD is less sharp).

The change of the $M_1$-$\Mmin$ scaling factor in massive halos can be
understood from the competition between accretion and destruction processes.
In general, massive halos accrete their satellites more recently than less 
massive halos. While the rate of satellite accretion for low mass halos 
peaks at $\sim 10$ Gyr ago, cluster-sized halos constantly accrete satellites 
until recently (\citealt{Zentner05}). As a consequence, there is less time 
for the orbit of satellites in a massive halo to decay through dynamical 
friction and for them to merge with the central galaxy to form a larger 
(brighter) central galaxy. In addition, the LRG samples we study are at 
redshift $\sim0.3$, which makes the accretion even more dominant. Therefore, 
the decrease in the ratio of $M_1$ and $\Mmin$ could be a manifestation of 
the favor of accretion over destruction in massive halos.

\begin{figure*}[t]
\epsscale{1.0}
\plotone{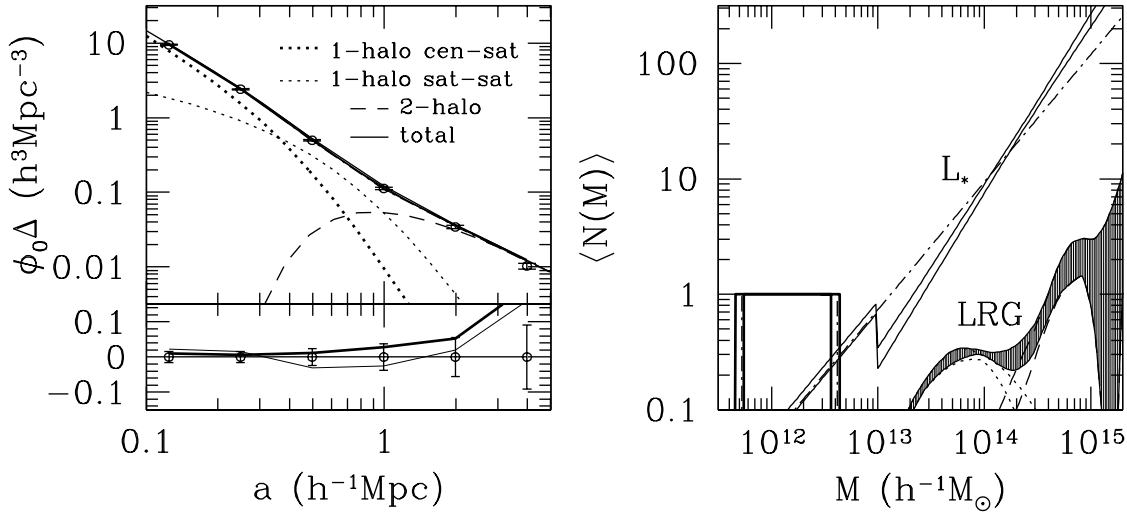}
\epsscale{1.0}
\caption[]{
\label{fig:crossNavg}
Mean occupation functions and cross-correlation functions of $L_*$ galaxies
and LRGs from the HOD modeling. {\it Left panels}: the predicted two-point
cross-correlation function between LRGs and $L_*$ galaxies separated into
contributions from central and satellite LRGs paired with $L_*$ galaxies within
same halos and LRGs paired with $L_*$ galaxies from different halos. The scale
$a$ is in units of proper $\hMpc$ to be consistent with that adopted
in \citet{Eisenstein05a}. The quantity $\phi_0=2.267\times 10^{-2} 
h^3{\rm Mpc}^{-3}$ is the proper number density of $L_*$ galaxies at 
the mean redshift $z\sim 0.3$. The lower panel shows the fractional difference
between model fits (thick for the flexible model and thin for the slope-fixed
model) and data (see the text).
{\it Right panel}: the mean occupation function for the $L_*$ galaxies in
massive halos (top solid curves) and that for the luminosity-bin LRGs. 
(bottom curves). The square window at the low mass end represents the mean 
occupation function of 
central
$L_*$ galaxies. The envelopes of the
mean occupation functions are derived from models with $\Delta\chi^2<4$. 
Dotted and dashed curves in the LRG mean occupation function are contributions
from central and satellite galaxies, respectively. The dot-dashed curve
shows the mean occupation function for the $L_*$ galaxies from a model with
the slope fixed to be 1.10 in the whole mass range.
}
\end{figure*}

\subsubsection{HOD for the Luminosity-Bin LRG Sample}

The cross-correlation between luminosity-bin LRG sample 
($-21.7<M_g<-21.2$) and the $L_*$ galaxies also leads to constraints on 
the HOD of the luminosity-bin LRG sample. The lower (solid) curves in 
the right panel of Figure~\ref{fig:crossNavg} are the $\Delta\chi^2<4$ 
envelope of the mean occupation function of the $-21.7<M_g<-21.2$ LRGs. 
The bump (dotted) around $\sim 10^{14}\hMsun$ is the contribution from 
central galaxies and the dashed curves represent contributions from 
satellites. The result shows that most of the LRGs in this luminosity bin
are central galaxies in halos of mass $\sim$2--20$\times 10^{13}\hMsun$, 
and a small fraction ($\sim$7\%) of them are satellites in more massive halos.

\subsection{Constraints on the HOD of $L_*$ Galaxies}

\begin{figure*}[t]
\epsscale{1.0}
\plotone{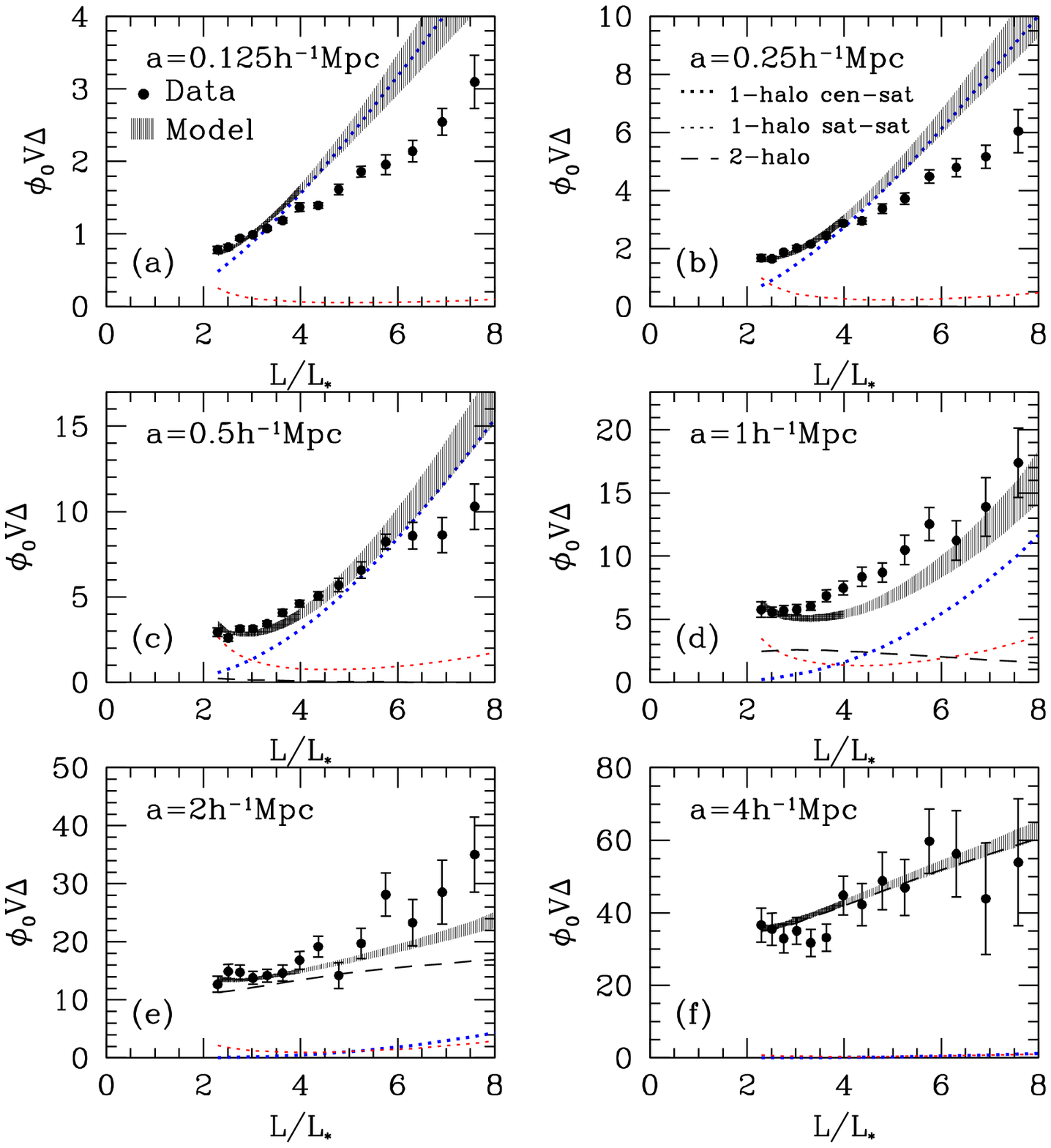}
\epsscale{1.0}
\caption[]{
\label{fig:crossL}
The luminosity dependence of the LRG--$L_*$ cross-correlation functions
at different scales. The six panels correspond to (proper) scales $a$=
0.125 to 4$\hMpc$, respectively, as labeled in each panel. The quantity 
$\phi_0=2.267\times 10^{-2} h^3{\rm Mpc}^{-3}$ is the proper number 
density of $L_*$ galaxies at the mean redshift $z\sim 0.3$, and 
$V=3(2\pi a^2)^{3/2}$ is the effective volume for the window function 
[see eq.~(\ref{eqn:window})].  In each panel, the shaded regions are
predictions from the modeling results. Since the model predictions are based 
on clustering information up to only $\sim 4L_*$, the shaded regions
below and above $4L_*$ are model interpolations and extrapolations,
respectively. The data points with error bars are the measurements 
in \citet{Eisenstein05a}. Thick (thin) dotted curves are contributions 
from central (satellite) LRGs paired with $L_*$ galaxies within common halos
(calculated from the best-fit HOD model), and dashed curves, which can be 
clearly seen in panels ({\it d})--({\it f}), 
represent the two-halo pair contribution. 
}
\end{figure*}

\citet{Zehavi05b} perform HOD modeling of the two-point auto-correlation 
functions of the MAIN galaxy sample and show that the two-point function 
can impose important constraints on the HOD of a sample of galaxies. 
In general, the mean occupation function is tightly constrained around 
$\NavgM\sim$ a few. It becomes loosely constrained towards higher halo 
masses because the two-point correlation function is less sensitive to 
the occupation distribution in these halos as a result of the steep drop 
of the halo mass function. Therefore, the analyses of the MAIN 
galaxy sample with low luminosity thresholds in \citet{Zehavi05b} cannot 
quite reveal how these low luminosity galaxies reside in massive halos. Since 
LRGs automatically pick out the massive halos, the cross-correlation between 
MAIN sample galaxies and LRGs provides us a nice way to study the halo 
occupation of MAIN sample galaxies in massive halos, enabling us to
better constrain the HOD of $L_*$ galaxies in massive halos.

Figure~\ref{fig:crosspar}{\it d} shows constraints on the concentration 
parameter $c_0$ and the high mass slope $\alpha_h$ of the mean occupation 
function for $L_*$ galaxies, marginalized over the other parameters. These two 
parameters are correlated in a sense that a higher $c_0$ corresponds to a 
higher $\alpha_h$. Higher $\alpha_h$ means that more $L_*$ galaxies reside 
in higher mass halos with lower concentrations and larger virial radius, 
so to maintain the small-scale (cross-)clustering the distribution 
of galaxies need to be more concentrated, i.e., a higher $c_0$. The high 
mass slope of the mean occupation function of $L_*$ galaxies is 
$1.49\pm 0.09$ (1-$\sigma$ range), steeper than the value 1.10 at lower 
mass. The concentration parameter $c_0$ represents the value extrapolated to 
$M_{\rm nl}\sim 1.5\times 10^{12}\hMsun$ according to 
$c=c_0(M/M_{\rm nl})^{-0.13}$.
Its 1-$\sigma$ range is found to be $9.4\pm 1.7$, which translates to 
$5.4\pm 1.0$ and $4.0\pm 0.7$ in halos of $10^{14}\hMsun$ and $10^{15}\hMsun$,
respectively, implying that the distribution of $L_*$ galaxies more or less
follows that of the dark matter, at radii the data can probe 
($\gtrsim 0.2\hMpc$ comoving). The mean occupation function of 
$L_*$ galaxies in massive halos is shown as upper (solid) curves in the 
right panel of Figure~\ref{fig:crossNavg}. These two curves are the envelope
determined by $\Delta\chi^2<4$. On average, about ten $L_*$ galaxies 
are expected to reside in a halo of mass $\sim 10^{14}\hMsun$. 

Our fits show that the
mean occupation function of $L_*$ galaxies at high halo masses becomes
steeper than the slope 1.10 at intermediate masses (and lower redshift).
We have also run a model with fixed slope 1.10 for the satellite 
mean occupation function in the whole mass range.
The fits to the 
LRG auto-correlation function and the LRG-$L_*$ cross-correlation function 
become much worse, 
with the overall $\chi^2$ increasing by $\sim$50. 
The resulting best fit to the cross-correlation function 
is plotted in the left panels of Figure~\ref{fig:crossNavg}. The lower panel
shows the fractional difference between the fits (thick curve for the flexible 
model and thin curve for the one with fixed slope) and the data.
The fit from the model with fixed slope is almost on top of the one from the 
more flexible model, but the error bars in the measurement are 
small and the change in $\chi^2$ is substantial (an increase of
$\sim10$ in the $\chi^2$ contributed by the cross-correlation function).
The corresponding high-mass 
end of the 
mean occupation function of $L_*$ galaxies from this more restricted model is 
shown as the dot-dashed line in the right panel of Figure~\ref{fig:crossNavg}. 
The single-slope model has
more $L_*$ satellites in halos less massive than 
$10^{14}\hMsun$, and the best-fit concentration parameter for $L_*$ 
galaxies is about a factor of two smaller; these two effects compensate each 
other to approximately maintain the amplitude of the small-scale 
cross-correlation. Unless the error bars in the measurements were 
underestimated (e.g., by a factor of two), the restricted model is highly 
disfavored by its much worse fits to the data ($\Delta\chi^2\sim 50$). 

The cause and implication of 
the steep inferred high mass slope of the $L_*$ galaxy occupation function 
is not clear. It may be related to the selection of $L_*$ galaxies ---
the sample used here is composed of galaxies in a bin of 1.5 magnitude around 
a redshift-dependent reference magnitude that is supposed to match 
$L_*$ at $z\sim 0.3$ \citep{Eisenstein05a}. Since the redshift range 
($0.16<z<0.44$) is not narrow, the $L_*$ sample should be regarded as an 
effective sample, rather than a uniform sample. It may also be caused by
the imperfection in the analytic model of the two-point cross-correlation
function. The small measurement errors in the two-point cross-correlation
function may require a more accurate model than the one used in this paper,
and allowing for uncertainty in the model accuracy
would increase the allowed range of the high mass slope.
Finally, it may also be related to the assumption that the occupation numbers 
of LRGs and $L_*$ galaxies in the same halo are uncorrelated (\S~3.2). Since
the effect of any correlation between the two occupation numbers becomes 
smaller as 
halo mass increases \citep{Simon09}, an anti-correlation between the LRG and 
$L_*$ occupation numbers would lead to a decrease in the high mass slope of 
the $L_*$ galaxy to conserve the number of LRG--$L_*$ pairs.
As a whole, we therefore have higher confidence in the value of $\langle N\rangle \sim 10$
at $M \sim 10^{14}\hMsun$, where the fits from the two models
cross, than we have in the slope $\alpha_h$ itself.

The modeling result also shows (see the left panel of 
Figure~\ref{fig:crossNavg}) that the LRG--$L_*$ cross-correlation function
is dominated by central LRGs paired with satellite $L_*$ galaxies on scales
less than $\sim 0.5 \hMpc$ (comoving), while above this scale the signal from 
satellite LRGs paired with satellite $L_*$ galaxies takes over until 
$\sim 1.5\hMpc$ (comoving), where the two-halo pairs start to dominate. 
We show below that the variation with scales in the contributing 
components is the key to understanding the scale dependence of the 
luminosity-dependent cross-clustering.
In the one-halo term, the signal from central LRGs mainly comes from
halos of mass a few times $10^{13}$ to $10^{14}\hMsun$, while the signal
from satellite LRGs is from halos of $\sim 10^{14.5}\hMsun$, where the LRG 
occupation number is a few. Although the occupation numbers of
LRGs and $L_*$ galaxies keep rising toward higher halo mass, more massive 
halos are too rare to make a significant contribution to the cross-correlation 
signal. 

\subsection{On the Scale Dependence of the Luminosity-Dependent 
Cross-Clustering}

For the LRG--$L_*$ cross-correlation, \citet{Eisenstein05a} find a strong 
dependence on LRG luminosity. The clustering amplitude becomes higher for
more luminous LRGs and varies by a factor of up to four over a factor of four 
in LRG luminosity (see their Fig.~2). Furthermore, the clustering amplitude 
increases more strongly with luminosity at smaller scales. We now show that 
these complex trends can be largely explained by the HOD results described 
above. 
We note that the HOD results are based on LRGs with luminosity $L<4L_*$,
and while the LRG--$L_*$ cross-correlation in \citet{Eisenstein05a} is 
measured up to an LRG luminosity of $8L_*$, we focus on the $L<4L_*$ results.
We emphasize that we do not intend to explain the data points for $L>4L_*$,
for which we extrapolate our results. 
Our purpose here is to give a qualitative interpretation of the scale-dependent
luminosity dependence of the cross-correlation between LRGs and $L_*$ galaxies.

From our modeling results, the HOD for LRGs in narrow luminosity bins can be 
readily constructed, similar to what we do for the $-21.7<M_g<-21.2$ sample. 
We do not refit auto-correlation functions of different LRG samples; rather, 
we apply the value of $\sigma_{M_g}$ and the scaling $L_c\propto M^p$ derived
from the faint and bright luminosity-threshold samples and interpolate (or
extrapolate) the satellite occupation function from these samples. 
Figure~\ref{fig:crossL} compares the predicted dependence of the 
cross-correlation on LRG luminosity to the observed one. The plotted
quantity $\phi_0 V\Delta$ is the average excess number of $L_*$ galaxies
around an LRG in an effective spherical volume of $V$ 
(see eq.~[\ref{eqn:Delta}]). We have already shown (Fig.~\ref{fig:crossNavg})
that our HOD model reproduces the overall scale dependence of the 
cross-correlation accurately. Figure~\ref{fig:crossL} shows that the model 
also captures the trend of stronger luminosity dependence at smaller scales.
Since the prediction is mostly based on modeling LRGs in the luminosity 
range of 2--4 $L_*$, it becomes less accurate at higher luminosities, where
it is essentially an extrapolation. 

Close inspection of the observed points shows that, at the smallest scales,
the clustering amplitude rises steeply and steadily from $\sim 2L_*$ to 
$\sim 8L_*$; while on larger scales, the luminosity dependence in the 
$2L_*$ -- $4L_*$ range is relatively flat, before steepening at higher
luminosities. In addition, the overall luminosity dependence is weaker
at larger scales. An intuitive understanding of these features can be built 
from our HOD modeling results. 

At very small scales, over all the LRG luminosity range, the cross-correlation
signal is dominated by {\it central} LRG paired with satellite $L_*$ galaxies. 
The thick dotted curves in Figure~\ref{fig:crossL} show this component from 
the model. The slope of this component is estimated as follows. 
Approximating the mean occupation function of central LRGs in a narrow 
luminosity bin as a Dirac-$\delta$ function, the two-point cross-correlation 
function at a separation $r$ between $L_*$ galaxies and LRGs is simply 
proportional to the pair number $\langle N_*(M) \rangle f(r;M)$, where 
$\langle N_*(M) \rangle$ is the mean occupation function of $L_*$ galaxies
in halos of mass $M$ and $f(r;M)$ is the fraction of $L_*$ galaxies located 
at a radius $r$ from the central LRG in halos of mass $M$. The function $f$ 
is just the distribution profile of $L_*$ galaxies, $f(r;M)=\rho(r;M)r^2/\int 
\rho(r;M)r^2 dr$. Using a power law to approximate the inner profile, 
$\rho(r;M)\propto (r/R_{\rm vir})^\gamma $, we have $f(r;M)\propto 
R_{\rm vir}^{-(3+\gamma)} \propto M^{-(1+\gamma/3)}$, where $R_{\rm vir}$ 
is the virial radius of the halo. Since $\langle N_*(M) \rangle \propto 
M^{\alpha_h}$, we see that the cross-correlation amplitude $\Delta \propto
M^{\alpha_h-(1+\gamma/3)}$. Noting that the central luminosity $L \propto
M^p$, the cross-correlation amplitude has the dependence on 
luminosity as $\Delta \propto L^{(\alpha_h-1-\gamma/3)/p}$. Inserting
typical values of the model results, $\alpha_h=1.49$, $\gamma=-1$ (inner NFW 
profile), and $p=0.66$ to the expression, we obtain $\Delta \propto L^{1.2}$, 
close to the observational result, which is roughly $\Delta \propto L^{1.1}$.

On larger scales (but still within the regime of one-halo pair domination), 
the contribution from {\it satellite} LRGs paired with satellite $L_*$ 
galaxies starts to show up at the low luminosity end, as can be seen at 
scales 0.25 and 0.5 $\hMpc$ in Figure~\ref{fig:crossL}. At a {\it fixed} scale, 
most satellite-satellite pairs come from halos of a narrow mass range.
As the luminosity in consideration increases, the occupation number of 
satellite LRGs at this halo mass decreases, leading to a decreasing 
contribution to the clustering amplitude from satellite-satellite pairs.
The opposing dependences of clustering contributions from central LRGs and 
satellite LRGs on luminosity flattens the overall luminosity dependence at the
low luminosity end, a feature seen in the observed clustering.

On much larger scales (e.g., 4$\hMpc$), the two-halo pairs dominate the 
cross-correlation between LRGs and $L_*$ galaxies, and the signal is 
proportional to the large-scale bias factor of LRGs. If the LRG bias factor 
is approximated by the halo bias factor, the luminosity-dependent clustering 
simply reflects the dependence of the halo bias factor on the halo mass. 
Around $M\sim 10^{14}\hMsun$, the halo bias factor can be approximated by 
a power law with index $\alpha_b\sim$ 0.3--0.4. This, together with the 
relation between central luminosity and halo mass $L\propto M^p$ 
($p\sim 0.66$), gives a luminosity dependence of the cross-clustering 
following roughly as $L^{0.5}$, which agrees well with the observed trend.

In modeling the LRG-$L_*$ cross-correlation, we make the assumption that
there is no correlation between the occupation numbers of LRGs and $L_*$ 
galaxies inside the same halo. Loosing this assumption would lead to 
changes in the contributions from one-halo central-satellite and 
satellite-satellite pairs. These changes would be at the level of fine 
details, and the above picture of the interplay among the three components
for interpreting the scale-dependent luminosity dependence of the 
cross-correlation would remain valid.

\section{Summary and Discussion}
\label{sec:discuss}

We have modeled the two-point auto-correlation functions 
of LRGs and the two-point cross-correlation functions between LRGs and $L_*$ 
galaxies in the SDSS, within the HOD framework, obtaining results on the mean
relation between central LRG luminosity and halo mass, the dispersion about 
this relation, the slope and amplitude of the satellite occupation function,
and the abundance of $L_*$ galaxies in massive halos.

The continuous rise toward small scales of two-point auto-correlation 
functions of LRGs implies that not all LRGs are the bright central galaxies 
in galaxy groups or clusters; a fraction of LRGs must be satellites 
to produce small scale, one-halo pairs. However, the satellite fraction is 
small and decreases with the LRG 
luminosity, e.g., $\sim$5--6\% for $M_g<-21.2$ and $\sim$2--3\% for 
$M_g<-21.8$ based on the HOD modeling. 
The characteristic minimum host halo mass of central LRGs 
($\Mmin$, at which 50\% of halos host a galaxy above the luminosity
threshold) 
is a few times $10^{13}\hMsun$ and increases with LRG luminosity.

\citeauthor{Zehavi05b} (\citeyear{Zehavi05b};
see also \citealt{Zheng07}) found a ratio 
$M_1/\Mmin\sim 20$ between the halo mass required to host a satellite above
a luminosity threshold and the mass required to host a central galaxy above
the same threshold. For these LRG samples, which populate higher mass halos
and have a median redshift $z\sim 0.3$, we find a smaller ratio, 
$M_1/\Mmin\sim 10$. A similar drop is seen for the brightest sample 
($M_r<-22$) in \citet{Zehavi05b}, which has a mean redshift $\sim0.16$. 
The decrease of the scaling factor reflects the balance between accretion 
and destruction 
of satellites (\citealt{Zentner05}) --- massive halos assemble more recently 
and their satellites have less time to merge with the central galaxy. The 
relatively higher redshift of the samples further strengthens this effect. 

The HOD of LRGs has been inferred using different methods in several recent 
investigations. To compare our results with others, one needs to pay 
attention to 
the differences in the sample definition, the underlying assumptions, and 
the modeling details.  \citet{Mandelbaum06} present a mass determination of 
host halos for two SDSS LRG samples based on galaxy-galaxy lensing 
measurements. The construction of their LRG samples is not identical to 
ours, but the halo masses determined from their two samples appear to 
be consistent with those of the two luminosity-threshold LRG samples we 
model. The galaxy lensing directly measures the halo masses, while our
results come from the mass distribution of halos and the galaxy assignment 
required to reproduce the observed clustering. The agreement between the
two results is therefore encouraging.
\citet{Wake08} model the projected 
two-point correlation functions for LRGs in the 2dF-SDSS LRG and QSO survey
with a three-parameter model. For the SDSS $z=0.21$ LRG sample, their
inferred mean occupation function is in general agreements with ours.
\citet{Blake08} perform HOD modeling of the two-point angular 
correlation function of $0.4<z<0.7$ SDSS LRGs with photometric redshifts. 
Their parameterization is a slight variation of our five-parameter model 
presented in Appendix~\ref{sec:appendixB}. For samples with similar number 
densities, their inferred halo mass scales, cutoff widths of central 
galaxy occupation function, and high mass slopes of satellite occupation 
function are close to what we obtain. 
\citet{Kulkarni07} constrain the HOD 
of SDSS LRGs with redshift-space two-point and three-point correlation 
functions, by comparing the measurements to those from mock catalogs 
generated through populating halos identified in $N$-body simulations. They 
use a three-parameter HOD description and assume no velocity bias. 
They find that redshift-space three-point correlation functions  
favor a lower high mass slope ($\sim 1.4$) for the satellite occupation 
function. Since the $z=0$ outputs of $\sigma_8=0.9$ simulations are used
in their modeling while the median redshift of LRGs is about 0.3, the 
effective $\sigma_8$ in their modeling is about 1.05. From 
Appendix~\ref{sec:appendixB} (eq.~[B3]), we see that the high mass slope 
from our modeling is 1.5 for such a high $\sigma_8$, which is close to the 
value favored by \citet{Kulkarni07}. However, we note that they use quite 
a different halo definition than ours (with a much lower overdensity
threshold), which complicates the comparison. A more detailed comparison
of the LRG modeling results with different methods and samples can be found
in \citet{Brown08}.

The inferred high mass slopes of the LRG occupation functions tend to 
be substantially larger than unity, either from our results or others (e.g.,
\citealt{Blake08,Kulkarni07,Wake08}). This appears to differ from observational 
inferences and theoretical predictions for low luminosity
samples (e.g., \citealt{Zehavi05b,Zheng07,Kravtsov04,Zheng05,Conroy06}). 
Is the steep slope of the inferred LRG occupation function a true feature 
or merely a result of modeling imperfections?
First of all, the steep slope should not be a result of any restriction in 
our HOD parameterizations. Our five-parameter HOD model 
(Appendix~\ref{sec:appendixB}) introduces a cutoff in $\Nsat$ at the low 
mass end so that the connection between the high mass end slope and that at 
$\Nsat\sim$ a few is broken. For the fainter LRG sample, the constraint on 
the slope up to $M\sim 2\times 10^{15}\hMsun$ remains tight even with our 
more flexible 
parameterization [see Fig.~\ref{fig:wp_LRG}({\it b})]. In the LRG
survey volume, there are not many halos that are more massive than 
$10^{15}\hMsun$. The true halo mass function in this volume can therefore 
deviate from the theoretical one used in the modeling. In 
Appendix~\ref{sec:appendixC}, we show that this fluctuation in the halo mass 
function does not seem to introduce any systematic biases in model 
fitting as it is already reflected in the covariance matrix of the data. 
To determine which features of the data drive the steep slope, we fit the 
faint LRG sample with the five-parameter model after excluding some data 
points and find that the two data points at $r_p\sim 1.7\hMpc$ and 
$2.7\hMpc$ play a large role 
--- if they are excluded, the slope drops from 1.8 to 1.6 (for $\sigma_8=0.8$). 
These scales are in the one-halo to two-halo transition region where the 
model is sensitive to the treatments of halo exclusion and scale-dependent 
halo bias. A more accurate scale-dependent halo bias with halos defined 
by spherical over-density (SO) could lead to a somewhat lower value of the high 
mass end slope (J.~L. Tinker, private communication). \citet{Reid09}
constrain the LRG HOD with the counts-in-cylinders multiplicity function
and the correlation function through populating halos in a 
simulation. They obtain a good fit by using SO halos and the high-mass end 
slope is found to be close to unity. Compared to friends-of-friends (FoF; 
\citealt{Davis85}) finder, the SO halo finder does not have the problem of 
linking two halos by a thin bridge of particles. We plan to pursue 
analytic models of galaxy clustering based on SO halo properties 
\citep{Tinker08} in future work.

Because of the steep high mass slope, our model fits predict that massive
clusters should host multiple LRGs. For example, the LRG occupation number 
of a $2\times 10^{15}\hMsun$ cluster would be about ten. Such a prediction 
can be tested with a cluster catalog if the mass can be determined. Using 
a sample of X-ray selected galaxy clusters at $0.2<z<0.6$, \cite{Ho07}
assess cluster membership for LRGs based on their photometric redshifts
and assign halo mass based on X-ray luminosity. They define halos as 
objects with mean density of 200 times
the critical density rather than the mean
background density as we do, and they 
assume $\Omega_m=0.238$. After correcting the 
differences in the halo definition and cosmology, our HOD result for the 
faint LRG sample matches theirs in the regime of $\NavgM \sim$ a few.
(However,
near the cutoff mass they find a much larger occupation number, while we have 
$\NavgM <1$.) In their catalog, there are only two very massive clusters with 
masses of $\sim 8\times10^{14}\hMsun$ and $\sim 1.1\times10^{15}\hMsun$ 
(corrected to be consistent with our halo definition), and the (corrected)
numbers of LRG members are about 8 and 14, respectively. We also performed
a rough calculation to associate LRGs in the faint sample to MaxBCG 
clusters \citep{Koester07} in the overlapped sky region and redshift 
range. The mass of each cluster is estimated from the total number of 
MAIN galaxies inside the virial radius, with calibration by weak lensing 
\citep{Sheldon07}. For each cluster, we infer the halo radius and velocity 
dispersion from the halo mass corrected to match our halo definition. 
LRGs that fall in the projected halo radius and three times the velocity 
dispersion along the line-of-sight direction are assigned as cluster 
members (with no completeness and edge correction applied). 
The number of LRG members for clusters more massive than 
$6\times 10^{14}\hMsun$ is found to range from 0 to 7. The predicted number
of LRGs in high mass halos (Fig.~\ref{fig:wp_LRG}) appears approximately 
consistent with the estimates from clusters. However, given the 
uncertainties in the mass estimator and the small number 
statistics, more work is needed to test our derived values of the high 
mass slope.

\begin{figure*}
\epsscale{1.0}
\plotone{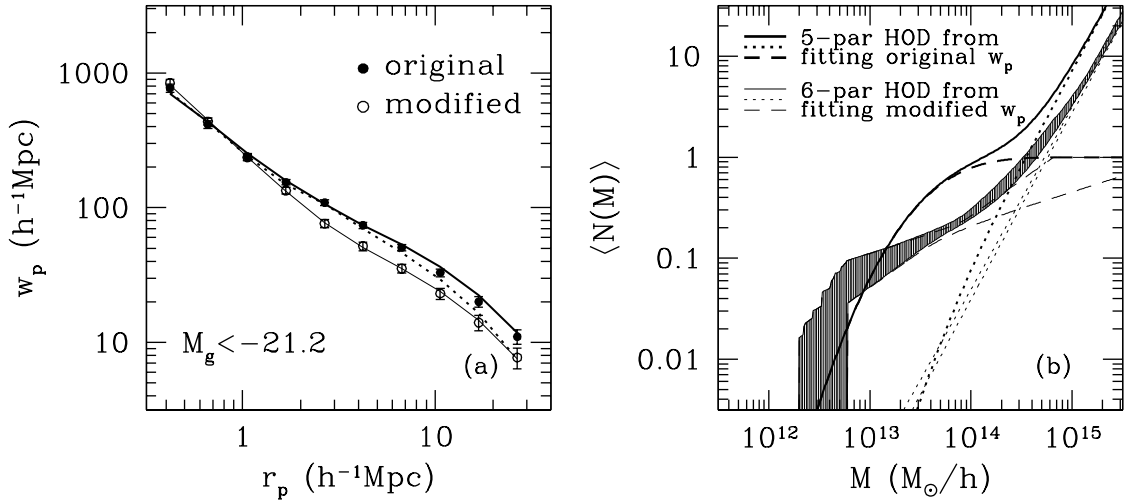}
\epsscale{1.0}
\caption[]{
\label{fig:samcmp}
Modified projected two-point correlation function and six-parameter HOD
fit. Panel ($a$) shows the original $w_p(r_p)$ data points (filled circles)
and the modified ones (open circles), which approximate those predicted by 
the semi-analytic model of \citet{Bower06} (see \citealt{Almeida08}). 
The thick curve is the best fit to the original data from the five-parameter
HOD model and the dotted curve is that from the HOD model presented in \S~3.1.
The thin curve is the best fit to the modified data from the six-parameter
HOD model (see the text).
The two curves are best HOD model fits. Panel ($b$) shows the best fit mean
occupation functions. The thick curves are from fitting the original $w_p(r_p)$
with the five-parameter model and the thin ones are the $\Delta\chi^2<1$
envelopes of mean occupation functions from fitting the modified $w_p(r_p)$
with a six-parameter model (see the text) and and those for all galaxies are 
shaded. Dashed, dotted, and solid curves
are for central, satellite, and all galaxies, respectively.
}
\end{figure*}

By modeling clustering of LRGs with different luminosities, we infer how
the mean luminosity of central galaxies changes with the mass of their host 
halos, $L_c\propto M^p$ with $p\sim$ 0.46--0.51 for $M\sim 10^{14}\hMsun$. 
This is consistent with the relation $M\propto L^2$ found by 
\citet{Mandelbaum06} based on galaxy-galaxy lensing measurements.
\citet{Padmanabhan09} model the clustering of a sample of photometrically
selected SDSS LRGs that are fainter than those in the spectroscopic samples 
we use. 
Combining their HOD modeling results with ours for LRGs at $z\sim 0.3$, 
we find that the value of $p$ is in the range of 0.4--0.5 over a larger 
range of LRG luminosity.
From a study of the halo occupation statistics of galaxies using galaxy
groups identified in the Two Degree Field Galaxy Redshift Survey (2dFGRS; 
\citealt{Colless01}), \citet{Yang05} infer a $L_c$--$M$ relation that is 
well described by a broken power law with $p\sim$ 2/3 and 1/4 below and 
above $10^{13}\hMsun$, respectively. The inferred $L_c$--$M$ relation by 
\citet{Vale04} from matching the luminosity function from 
the 2dFGRS with the theoretical subhalo mass function has a similar 
high-mass slope $p\sim 0.28$. \citet{More09} constrain the high-mass slope 
to be $0.28_{-0.09}^{+0.07}$ for $z<0.072$ SDSS galaxies based on satellite 
kinematics. The high-mass slope extrapolated from the HOD modeling results 
of SDSS MAIN galaxies \citep{Zheng07} is also about 0.3.
The value we obtain at the high-mass end from modeling the LRG clustering 
differs significantly from these results. It is interesting to see whether 
such a difference can be explained by the difference in the galaxy samples. 
The 2dFGRS luminosity is in the $b_J$ band, while we use the SDSS $g$-band 
luminosity. Since the wavelength coverages of these two bands are close, 
it seems unlikely that the band difference can cause the apparent 
discrepancy. The other 
thing to notice is that the 2dFGRS galaxies and the SDSS MAIN galaxies have 
a mean redshift 
$\sim 0.1$ and the LRG galaxies in our analysis are located around redshift 
0.3. Could the discrepancy indicate an evolution effect over the intervening
$\sim$2 
billion years? Through fitting restframe $B$-band galaxy luminosity functions 
at different redshifts using a CLF approach, \citet{Cooray05} finds that 
the data are compatible with a halo-mass-dependent central galaxy luminosity 
evolution, with the high mass slope $p$ increasing with redshift. The 
fitting results of \citet{Cooray05} imply that $p$ could be as high as 0.5 
at $z\sim 0.3$, close to our inferred value. 

If we take the inferred values of $p$ at $z\sim 0.3$ from our analysis and 
at $z\sim 0.1$ from the 2dFGRS studies at face value, they 
indicate that the luminosity evolution of central bright 
galaxies depends on halo mass in the sense that either galaxies in more 
massive halos fade more or the fraction of stars that were assembled into 
central galaxies from mergers between $z\sim 0.3$ and $z\sim0.1$ is smaller 
in more massive halos.  \citet{Bernardi06}'s study of the properties of 
early-type galaxies in the SDSS as a function of local environment and 
redshift suggests that star formation in early-type galaxies happens 
earlier in dense regions and lasts over a shorter time-scale,
which implies that central galaxies in more 
massive halos on average experience star formation at an earlier epoch. 
Since younger stellar populations fade faster than older ones, this seems 
to rule out the possibility that luminosity of central galaxies in more 
massive halos fades more between $z\sim 0.3$ and $z\sim 0.1$. We are 
left with the possibility that mergers in more massive halos are less 
efficient in adding stars to the central galaxies, which appears to be 
consistent with
our finding of the drop of $M_1/\Mmin$ and its interpretation
based on the competition between accretion and destruction as a function
of halo mass. This is also in line with the results of LRG evolution 
in \citet{Brown08} from HOD modeling of their clustering from 
$z\sim 0.2$ to $z\sim 1.0$ in the NOAO Deep Wide-Field Survey. 

Since LRGs naturally separate out massive halos, our HOD modeling of the 
two-point cross-correlation functions between LRGs and $L_*$ galaxies 
circumvents the usual challenge in constraining the HOD of $L_*$ galaxies in 
massive halos from their two-point auto-correlation function. We show that
the cross-correlation data are consistent with the case where the 
distribution of $L_*$ galaxies inside massive halos roughly follows that 
of the matter at distances greater than $\sim 0.2\hMpc$ and that the mean
number of $L_*$ galaxies scales with halo mass as $M^{1.5}$. The slope of the
mean occupation function is steeper than what is found for $z\sim 0$ galaxies,
which is $\sim 1.10$. There may not be inconsistency between the results.
The constraint for the slope for $z\sim 0$ galaxies from auto-correlation 
function is not sensitive to the occupation in very massive halos, while 
here from cross-correlation with LRGs, we have a better constraint on the 
$L_*$ occupation function in massive halos. On the other hand, the selection 
of $L_*$ galaxies and its redshift dependence make it hard to do a precise
comparison between results at $z\sim 0$ and at $z\sim 0.3$.

The luminosity 
dependence of the LRG--$L_*$ galaxy cross correlation depends on scale in
a rather complex way \citep{Eisenstein05a}. By separating 
contributions from pairs of $L_*$ galaxies with central and satellite LRGs 
in common halos and in different halos, our 
HOD modeling results explain these trends, in a relatively transparent way.
At a fixed scale, the luminosity dependence in the cross correlation reflects
the luminosity-dependent HOD of LRGs. As the scale in consideration changes,
the relative contributions of the one-halo central-satellite, the one-halo 
satellite-satellite, and the two-halo LRG-$L_*$ pairs to the cross-correlation 
function vary, which leads to the scale-dependent luminosity dependence of 
the cross correlation. 

Our LRG modeling results establish the relation between massive galaxies 
and dark matter halos at $z\sim 0.3$, which itself provides useful tests
to models of formation of massive galaxies. \citet{Almeida08} present 
predictions for properties of LRGs in semi-analytic galaxy formation models.
They show that the \citet{Bower06} model, which is based on the Millennium 
simulation \citep{Springel05}, predicts a $z=0.24$ LRG luminosity
function that is in good agreement with the observation, although it fails at
$z=0.5$. Compared to the measurement, the \citet{Bower06} model seems to 
predict the $z=0.24$ LRG two-point correlation function remarkably well on 
both small and large scales (see their Fig.13). However, the LRG HOD predicted
in this model differs significantly from our modeling results presented in this
paper: (a) in the \citet{Bower06} model, LRGs in the fainter sample can 
reside in halos of $10^{12}\hMsun$ (with mean occupation number of $\sim$0.01), 
much lower than the mass scale we infer; (b) the mass of halos that can on 
average host one LRG is about $3\times 10^{14}\hMsun$, higher than what we 
find (see Fig.11 in \citealt{Almeida08} and Fig.16 in \citealt{Wake08}); 
(c) the probability distribution of LRG host halo masses is broad
(see  Fig.16 in \citealt{Wake08}), ranging from $10^{12}\hMsun$ to 
$10^{15}\hMsun$, rather than narrowly peaked around a few times 
$10^{13}\hMsun$ as we find (Fig.~\ref{fig:wp_LRG}$d$). 

Does the discrepancy between the theoretically predicted HOD and our 
observationally inferred HOD imply that our HOD parameterization is not 
generic enough to model LRG clustering? To investigate the problem,
we modify the five-parameter HOD model (Appendix~\ref{sec:appendixB}) 
to mimic the shape of the mean occupation function predicted in 
\citet{Bower06} model. The parameterization for the satellite mean occupation 
function remains unchanged (with three parameters). For the central galaxy 
occupation function, we model it as a linear (in logarithmic space) ramp 
going from $\langle N_{\rm min} \rangle$ at $\Mmin$ to unity at $M_u$ and 
staying at unity for $M>M_u$. In total, this parameterization has six free 
parameters. With the measured $w_p(r_p)$, we find that the best-fit HOD 
from this six-parameter model (not shown in Figure~\ref{fig:samcmp}$b$) 
closely follows the result from the five-parameter model (shown as thick curves in Figure~\ref{fig:samcmp}$b$). Therefore,
change in the HOD parameterization does not solve the discrepancy and the 
five-parameter model is not inadequate in modeling LRG clustering. A close
look at Figure~13 in \citet{Almeida08} shows that the predicted two-point 
correlation function does not match the data perfectly --- it is $\sim$15\%
higher on scales of 0.1--2$\hMpc$ and $\sim 30\%$ lower on scales larger than
2$\hMpc$ (note that the vertical range of the plot is over eight orders of 
magnitude). We therefore modify the amplitude of the observed $w_p(r_p)$ data 
points to mimic the predicted correlation function and perform an HOD fit
with the six-parameter model (by adopting the predicted number density,
which is 10\% lower than the observed one). The results on the mean 
occupation functions are shown in panel ($b$) of Figure~\ref{fig:samcmp}
as thin curves. 
Note that the shaded region represents the $\Delta\chi^2<1$ envelopes
and the linear ramp for the central galaxy mean occupation number (thin dashed 
curves) has not reached unity at the highest mass in the plot.
The mean occupation function can extend to halos of mass
as low as $2\times 10^{12}\hMsun$ and reaches unity around 
$4\times 10^{14}\hMsun$, which appears to be approximately consistent with 
the prediction of the \citet{Bower06} model. From the above investigations, 
we conclude that the discrepancy 
between the theoretically predicted LRG HOD and the observationally inferred
LRG HOD reflects the imperfection of the semi-analytic galaxy formation 
model --- the $15-30\%$ discrepancies with observed clustering are
real and physically significant ---
rather than limitations in our HOD parameterization. 
The results suggest that the mechanism of turning blue galaxies to red in 
the semi-analytic model is too efficient in halos of a few times 
$10^{12}\hMsun$ and not efficient enough in more massive halos. 
Our HOD modeling
results thus provide important tests to galaxy formation theory.

In combination with passive evolution 
of LRGs and a halo merger history (e.g., \citealt{White07,Seo08}), our 
modeling results can be used to predict the HOD of LRGs and the clustering 
of LRGs at lower or slightly higher redshifts. Supplemented with corresponding 
observations at these redshifts, we would be able to test our understanding 
of the formation and evolution of massive galaxies. 
Constraints on the HOD of LRGs are also useful for cosmological parameter
investigations based on LRG clustering. For example, the most precise 
measurements of the large scale galaxy power spectrum have come from the 
SDSS LRG sample (\citealt{Tegmark06,Percival07}), and the principal 
limitation in interpreting these measurements is the uncertain level of
scale-dependent bias between galaxy and matter power spectra in the mildly
non-linear regime. With HOD constraints like those derived here, this
scale-dependent bias can be calculated and corrected \citep{Yoo09}.
As another example, HOD constraints could be combined with 
galaxy-galaxy lensing measurements of the LRG-mass cross-correlation
(R. Mandelbaum et al., in preparation) to improve determinations of 
$\sigma_8$ and $\Omega_m$ \citep{Yoo06}. The role of LRGs in cosmological
studies seems destined to grow with surveys that target large numbers of
LRGs to measure baryon acoustic oscillations, including AAOmega LRG survey
\citep{Ross08} and the Baryon Oscillation Spectroscopic Survey (BOSS), part of
a proposed successor to SDSS-II. Understanding the evolving relation between 
LRGs and dark matter halos will be crucial to exploiting their power as 
cosmological probes and to revealing the physics that governs the formation 
of the most massive galaxies in the universe.

\acknowledgments 
We are grateful to Jeremy Tinker and Jaiyul Yoo for valuable discussions. 
We thank Erin Sheldon for providing us with his cluster lensing results and 
acknowledge the Aspen Center for Physics, where discussions in a stimulating 
atmosphere led to a more complete presentation of this paper. We thank Tobias
Baldauf for pointing out typos in an early draft. Finally, we 
thank the referee for a careful reading of the paper and for helpful comments 
that improved the paper.

At an early stage of this work, ZZ was supported by NASA through Hubble 
Fellowship grants HF-01181.01-A, awarded by the Space Telescope Science 
Institute, which is operated by the Association of Universities for Research 
in Astronomy, Inc., for NASA, under contract NAS 5-26555. ZZ also 
acknowledges support from the Institute for Advanced Study through a John
Bahcall Fellowship. IZ and ZZ acknowledge support from NSF grant 
AST-0907947  and DW acknowledges support from NSF grant AST-0707985.
IZ was further supported by NASA through a contract issued by the Jet 
Propulsion Laboratory. YPJ is supported by NSFC (10533030), by the 
Knowledge Innovation Program of CAS (No. KJCX2-YW-T05), and by 973 Program
(No.2007CB815402).

Funding for the SDSS and SDSS-II has been provided by the Alfred P. Sloan
Foundation, the Participating Institutions, the National Science Foundation,
the U.S. Department of Energy, the National Aeronautics and Space
Administration, the Japanese Monbukagakusho, the Max Planck Society, and
the Higher Education Funding Council for England. The SDSS Web Site is
http://www.sdss.org/.

The SDSS is managed by the Astrophysical Research Consortium for the
Participating Institutions. The Participating Institutions are the
American Museum of Natural History, Astrophysical Institute Potsdam,
University of Basel, Cambridge University, Case Western Reserve
University, University of Chicago, Drexel University, Fermilab,
the Institute for Advanced Study, the Japan Participation Group,
Johns Hopkins University, the Joint Institute for Nuclear Astrophysics,
the Kavli Institute for Particle Astrophysics and Cosmology,
the Korean Scientist Group, the Chinese Academy of Sciences (LAMOST),
Los Alamos National Laboratory, the Max-Planck-Institute for Astronomy (MPIA),
the Max-Planck-Institute for Astrophysics (MPA), New Mexico State University,
Ohio State University, University of Pittsburgh, University of Portsmouth,
Princeton University, the United States Naval Observatory, and the
University of Washington.

\appendix

\section{Role of Halo Mass Scales on the Departures from a Power
 Law in the Galaxy Two-Point Correlation Function}
\label{sec:appendixA}

\begin{figure}[h]
\label{fig:appendixA}
\plotone{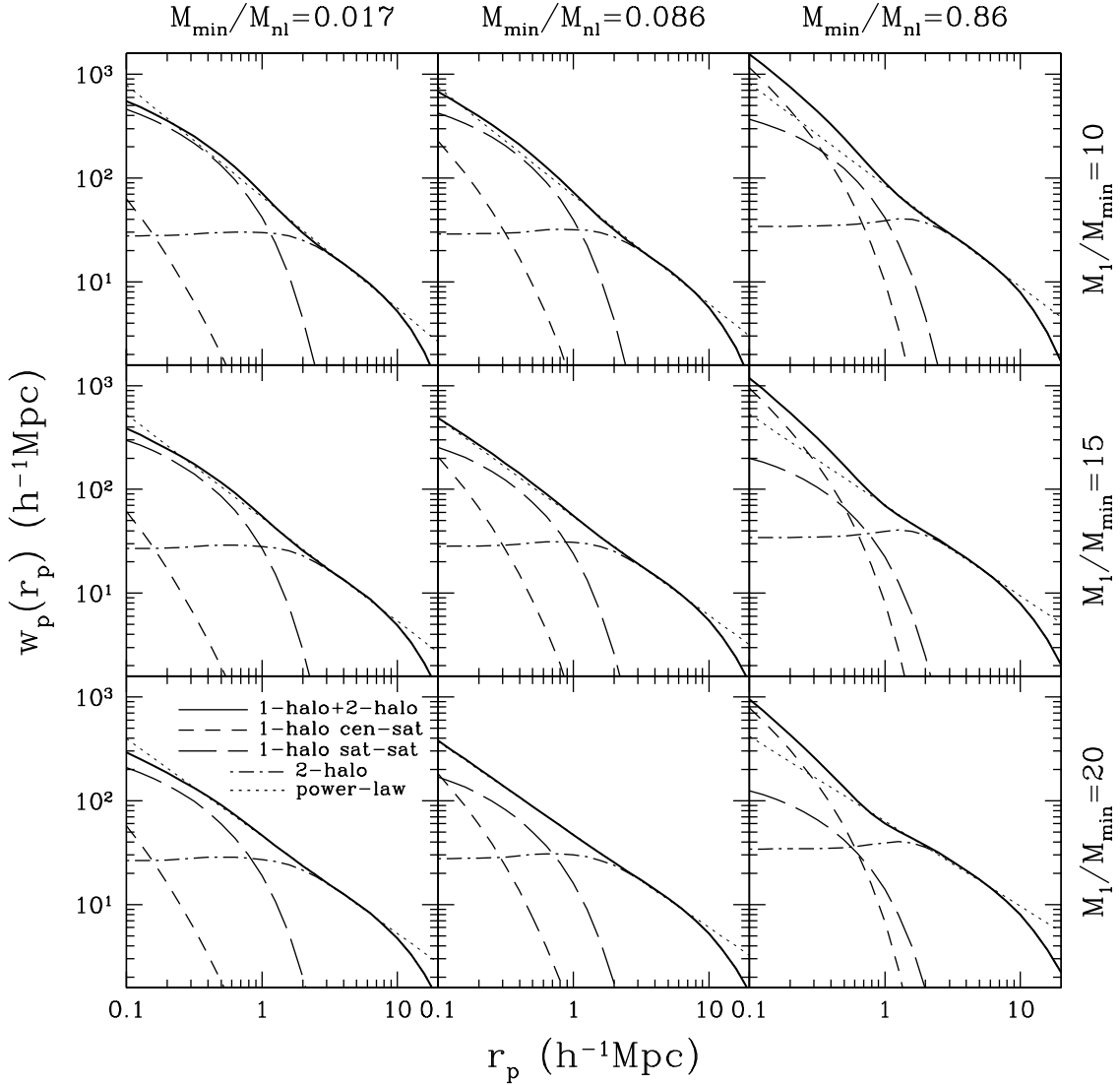}
\caption[]{Impact of the $M_1/\Mmin$ and $\Mmin/M_{\rm nl}$ ratios on 
departures from a 
power law in the galaxy two-point correlation function. Each row (column) 
has the same $M_1/\Mmin$ ($\Mmin/M_{\rm nl}$) with the value marked on the 
right (top) of the plot. In each panel, the solid line shows the predicted
projected correlation function, $w_p(r_p)$. The short and long dashed curves 
are the one-halo terms from central-satellite and satellite-satellite 
galaxy pairs, respectively, and the dot-dashed curve is the two-halo term.
The dotted line is a power-law fit to $w_p(r_p)$ in the range of 1--10$\hMpc$ 
to guide the eye. 
}
\end{figure}

Departures of the galaxy two-point correlation function from a pure
power law have been observed for both low and high redshift galaxies
(\citealt{Hawkins03,Zehavi04,Ouchi05,Coil06,Lee06}). In the two 
luminosity-threshold LRG samples studied in this paper, the departures are 
also clearly seen, shown as an upturn in the two-point correlation function 
at small scales (see also \citealt{Zehavi05a}).  Such departures have been 
nicely explained by the HOD 
model as the transition from a regime dominated by one-halo pairs on small 
scales to that dominated by two-halo pairs on large scales \citep{Zehavi04}.  
The strength of the departures of the galaxy two-point correlation function 
from a pure power law is closely related to the amplitude of the one-halo
term, which itself depends on the scatter in the occupation number and
the halo mass function (\citealt{Benson00,Berlind02}). It would be helpful 
to 
gain a better understanding of the key factor that determines the strength
of the departures.

For luminosity-threshold samples, there are two mass scales, the 
characteristic minimum mass $\Mmin$ of halos that host central 
galaxies above the luminosity threshold and the mass scale $M_1$
of halos that on average host one satellite galaxy above the 
luminosity threshold. An additional mass scale is the nonlinear mass 
$M_{\rm nl}$, marking a transition in the halo abundance from  
a power-law form to an exponential cutoff at high mass.
We can thus identify two ratios that can shape the one-halo term: $M_1/\Mmin$
and $\Mmin/M_{\rm nl}$. The  $M_1/\Mmin$ ratio affects the shape of the galaxy
occupation functions and tells us how quickly the transition 
from sub-Poisson to Poisson scatter occurs
when going to higher halo masses. A smaller $M_1/\Mmin$ increases the 
importance of one-halo
pairs by increasing the satellite fraction and thus results in a higher 
amplitude
of the one-halo term. The $\Mmin/M_{\rm nl}$ ratio determines which part of the 
halo mass function the galaxy sample probes. The slope of the one-halo term 
reflects the drop of the halo mass function toward high masses. For a galaxy 
sample that probes the exponential tail of the halo mass function 
($\Mmin/M_{\rm nl}\ga 1$), 
we expect a steep drop in the one-halo term and thus a sharp upturn around 
the one-halo to two-halo transition scale.

To figure out the relative importance of $M_1/\Mmin$ and $\Mmin/M_{\rm nl}$ 
on the
departures from a power law in the galaxy two point correlation function,
we calculate the predicted correlation functions on a $3\times 3$ grid
of $M_1/\Mmin$ and $\Mmin/M_{\rm nl}$, as shown in 
Figure~\ref{fig:appendixA}.
In each panel, the short dashed curve is the one-halo term from 
central-satellite pairs, the long dashed curve is the one-halo term from
satellite-satellite pairs, and the dot-dashed curve is the two-halo term.  
The dotted line is a power-law fit to $w_p(r_p)$ in the range of 1--10$\hMpc$ 
for comparison. The satellite-satellite pair contributions are similar at 
a fixed $M_1/\Mmin$, since $\Nsat$ is assumed to be proportional to halo mass.
In a massive halo of fixed mass, however, the contribution of 
central-satellite pairs relative to satellite-satellite pairs is larger for 
a higher $\Mmin/M_{\rm nl}$ sample because of its low satellite occupation 
number.  This leads to a larger slope across the central-satellite and 
satellite-satellite contributions for a higher $\Mmin/M_{\rm nl} $ sample, thus
a steeper upturn in the correlation 
function\footnote{We thank Alison Coil,
Jeremy Tinker, and Risa Wechsler for helpful discussions that led us
to separately investigate the central-satellite and satellite-satellite pair 
contributions.}. In more detail, the slope 
across the central-satellite and satellite-satellite contributions should 
depend on the $M_1$ mass scale, but $M_1/\Mmin$ should not vary by an 
extremely large factor at least for a threshold galaxy sample. So a dependence
on $M_1$ can be translated to that on $\Mmin/M_{\rm nl}$.
When probing the 
exponential tail of the halo mass function for observed galaxy samples,
we expect a steeper slope in both the central-satellite
and satellite-satellite one-halo terms. This in turn leads to a more
prominent rise at small scales.

From the results, we conclude that overall the $\Mmin/M_{\rm nl}$ ratio plays 
a much more important role than $M_1/\Mmin$ in driving departures from a 
power law in galaxy 
two-point correlation functions. That is, a steeper drop in the halo mass 
function is translated to a steeper radial cutoff in the one-halo term, 
leading to a stronger inflection where the one-halo and two-halo term join.
This explains why the departure is 
stronger for more luminous galaxy samples (e.g., \citealt{Zehavi05a,Zehavi05b}) that 
probe the high mass end of the halo mass function. This also explains why 
the departures are more prominent at higher redshifts (e.g., 
\citealt{Ouchi05}) --- the fast drop of the nonlinear mass $M_{\rm nl}$ 
toward high 
redshift makes the observed (bright) galaxies more likely to probe the 
exponential tail of the halo mass function.
A more detailed theoretical investigation of the departure from a power 
law in the two-point correlation function of galaxies and its dependence 
on galaxy properties will be presented elsewhere.

\section{LRG HODs with a Five-Parameter Model and Dependence on Cosmology}
\label{sec:appendixB}

\begin{figure}
\plotone{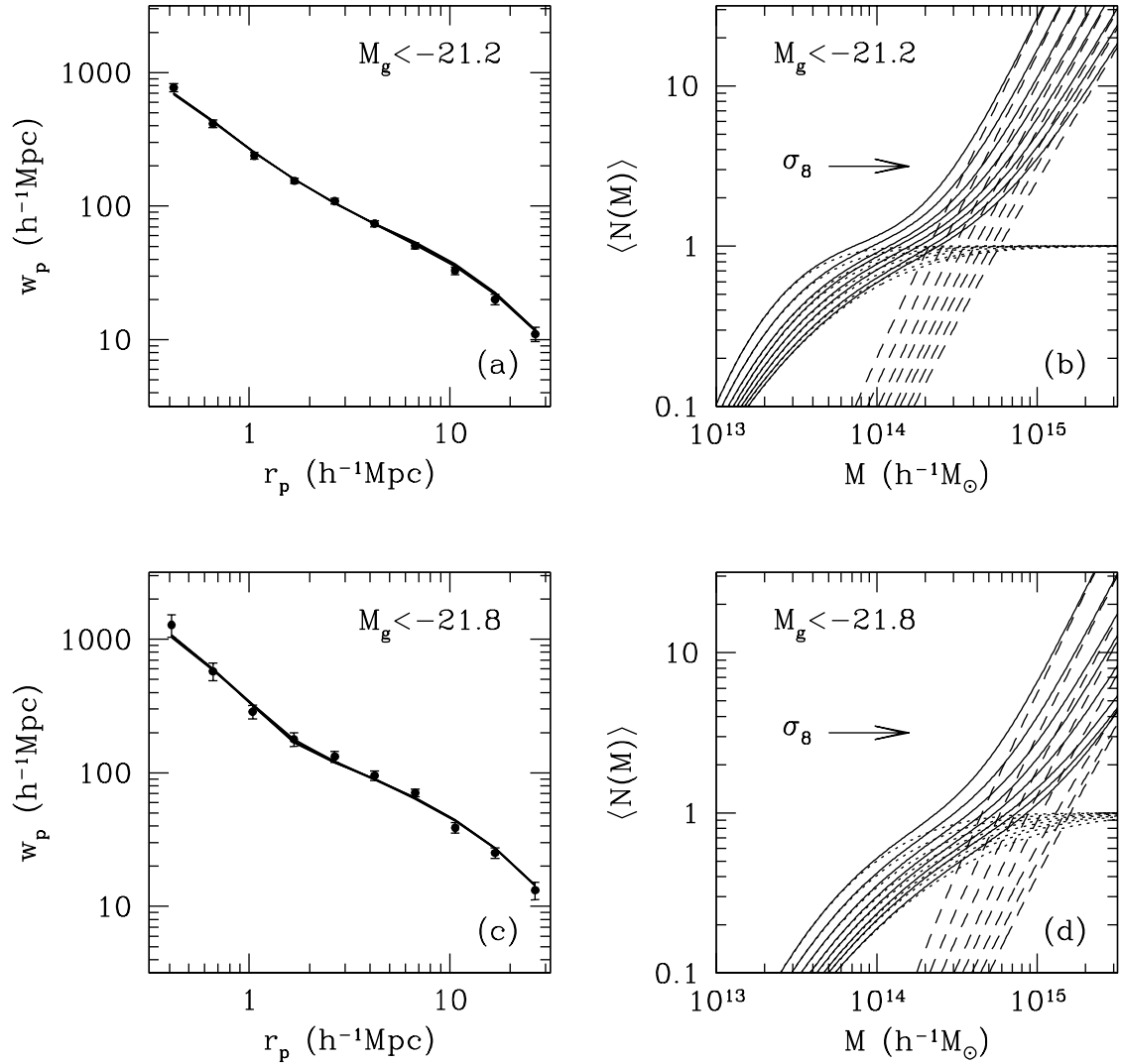}
\caption[]{
\label{fig:5par_sig8}
Dependence of the LRG HOD on cosmology. Panel ({\it a}) shows the measured
two-point correlation function (data points with error bars) of the 
$M_g<-21.2$ sample and the best HOD fits (solid curves) for cosmological 
models differing only in $\sigma_8$. Note that the best fits overlap with 
each other. Panel ({\it b}) shows
the mean occupation functions (solid curves) corresponding to the best fits,
separated into central (dotted) and satellite (dashed) contributions. From
left to right, the HODs correspond to $\sigma_8$ values increasing from 
0.65 to 1.00 with a step-size 0.05. The two bottom panels are similar, but 
for the $M_g<-21.8$ LRG sample.
}
\end{figure}

The LRG clustering modeling in the main text assumes a flexible HOD 
parameterization, which reveals details of the constraining power of 
the two-point correlation functions on the HOD. In general, the HODs for 
luminosity-threshold samples predicted by galaxy formation models  
can be well described by a less flexible (i.e., more restricted) parametric 
form. The mean occupation
function is well characterized by a step-like function for central
galaxies and a power law-like function for satellite galaxies 
(\citealt{Kravtsov04,Zheng05}). We present here the modeling results for 
the two luminosity-threshold LRG samples with an alternative, five-parameter 
HOD model and give their dependence on cosmology. This set of results would 
be useful for comparisons with other work and for making mock catalogs 
in a wide range of cosmological models.

The mean occupation function of a luminosity-threshold LRG sample,
being the sum of central and satellite mean occupation functions, is
parameterized as (see \citealt{Zheng05,Zheng07})
\begin{equation}
\langle N(M)\rangle=\frac{1}{2} 
\left[1+\erf\left(\frac{\log M-\log\Mmin}{\sigM}\right)\right]
\left[1+\left(\frac{M-M_0}{M_1^\prime}\right)^\alpha\right],
\end{equation}
where $\erf$ is the error function
\begin{equation}
\label{eqn:erf}
{\rm erf}(x)=\frac{2}{\sqrt{\pi}} \int_0^{\rm{x}} e^{-t^2} dt .
\end{equation}
The distribution of the occupation number of central galaxies and satellite
galaxies are assumed to follow the nearest-integer and Poisson distributions,
respectively, as usual. The five free parameters are the mass scale $\Mmin$ 
and width $\sigM$ of the cutoff profile for the mean occupation function of 
central galaxies and the cutoff mass scale $M_0$, normalization $M_1^\prime$,
and high mass slope $\alpha$ of the mean occupation function of
satellite galaxies.

We vary the normalization $\sigma_8$ (at $z=0$) of the matter fluctuation 
power spectrum from 0.65 to 1.00 with a step-size 0.05. Other 
cosmological parameters are assumed to be 
($\Omega_m$, $\Omega_\Lambda$, $\Omega_b$, $n_s$, $h$)=(0.24, 0.76, 0.04, 0.95, 0.73). 
We perform an MCMC run 
for each LRG sample under each cosmological model and obtain the marginalized 
distribution for each of the five HOD parameters. 

In general, for a higher $\sigma_8$, the nonlinear mass increases and there
are more high mass halos. To conserve the galaxy number density, the cutoff
mass scale $\Mmin$ needs to increase. The increase in $\Mmin$ turns out to be
slower than that in $M_{\rm nl}$, which leads to a lower halo bias that is
necessary to maintain the large scale clustering of galaxies (i.e., the 
square of $b_g\sigma_8$). The width $\sigM$ of the cutoff increases so that 
some LRGs are populated to lower mass halos to further adjust the large scale 
bias factor and the galaxy number density. The mean occupation function of 
satellites also
responds to the $\sigma_8$ change to match the small scale clustering 
amplitude: the mass scale $M_1^\prime$ increases and the slope $\alpha$ 
decreases with increasing $\sigma_8$. For a large range of $\sigma_8$, 
our HOD modeling yields almost identical best fits to the data points, 
similar to what is found and discussed in \citet{Zheng07a}.
 
While the dependence of 
$\log \Mmin$, $\sigM$, $\log M_0$, $\log M_1^\prime$ or $\alpha$ on 
$\sigma_8$ appears to be quite close to linear, we fit the results
by adding a quadratic term for a better accuracy. 
The HOD parameters for the $M_g<-21.2$ LRG sample are
\begin{eqnarray}
\log \Mmin      &=& 13.673 +1.419(\sigma_8-0.8)-1.706(\sigma_8-0.8)^2, \nonumber \\
\sigma_{\log M} &=& \,\,\,0.621  +0.908(\sigma_8-0.8)-0.935(\sigma_8-0.8)^2, \nonumber \\
\log M_0        &=& 12.339 +0.658(\sigma_8-0.8)+9.206(\sigma_8-0.8)^2, \nonumber \\
\log M_1^\prime &=& 14.533  +1.248(\sigma_8-0.8)-1.394(\sigma_8-0.8)^2, \nonumber \\
\alpha          &=& \,\,\,1.832  -1.326(\sigma_8-0.8)+0.523(\sigma_8-0.8)^2.
\end{eqnarray}
The typical 1--$\sigma$ uncertainties for the five HOD parameters are
0.06, 0.07, 0.6, 0.025, and 0.08, respectively.
The HOD parameters for the $M_g<-21.8$ LRG sample are
\begin{eqnarray}
\log \Mmin      &=& 14.304 +1.694(\sigma_8-0.8)\,\,\,-1.810(\sigma_8-0.8)^2, \nonumber \\
\sigma_{\log M} &=& \,\,\,0.797  +0.761(\sigma_8-0.8)\,\,\,-0.614(\sigma_8-0.8)^2, \nonumber \\
\log M_0        &=& 12.491 -1.476(\sigma_8-0.8)+29.983(\sigma_8-0.8)^2, \nonumber \\
\log M_1^\prime &=& 14.946  +1.616(\sigma_8-0.8)\,\,\,-0.712(\sigma_8-0.8)^2, \nonumber \\
\alpha          &=& \,\,\,1.717  -0.589(\sigma_8-0.8)\,\,\,-7.437(\sigma_8-0.8)^2.
\end{eqnarray}
The typical 1--$\sigma$ uncertainties for the these five HOD parameters are
0.06, 0.055, 0.7, 0.1, and 0.4, respectively.
All the masses are in units of $\hMsun$.

For results with cosmological models with $\Omega_m$ different than 0.24, one 
only needs to change the three mass scales by a factor of $\Omega_m/0.24$ 
(see \citealt{Zheng02,Zheng07a}). This scaling assumes that the {\it shape}
of the linear power spectrum stays fixed, with a change in $h$ or $n_s$ 
compensating the impact of changing $\Omega_m$. Therefore, the solutions 
here cover a wide range of cosmological models in the ($\Omega_m$,$\sigma_8$) 
plane.

\section{Halo Number Fluctuation in the LRG Survey Volume and the Covariance 
Matrix}
\label{sec:appendixC}

\begin{figure}[h]
\label{fig:mf_fluctuation}
\plottwo{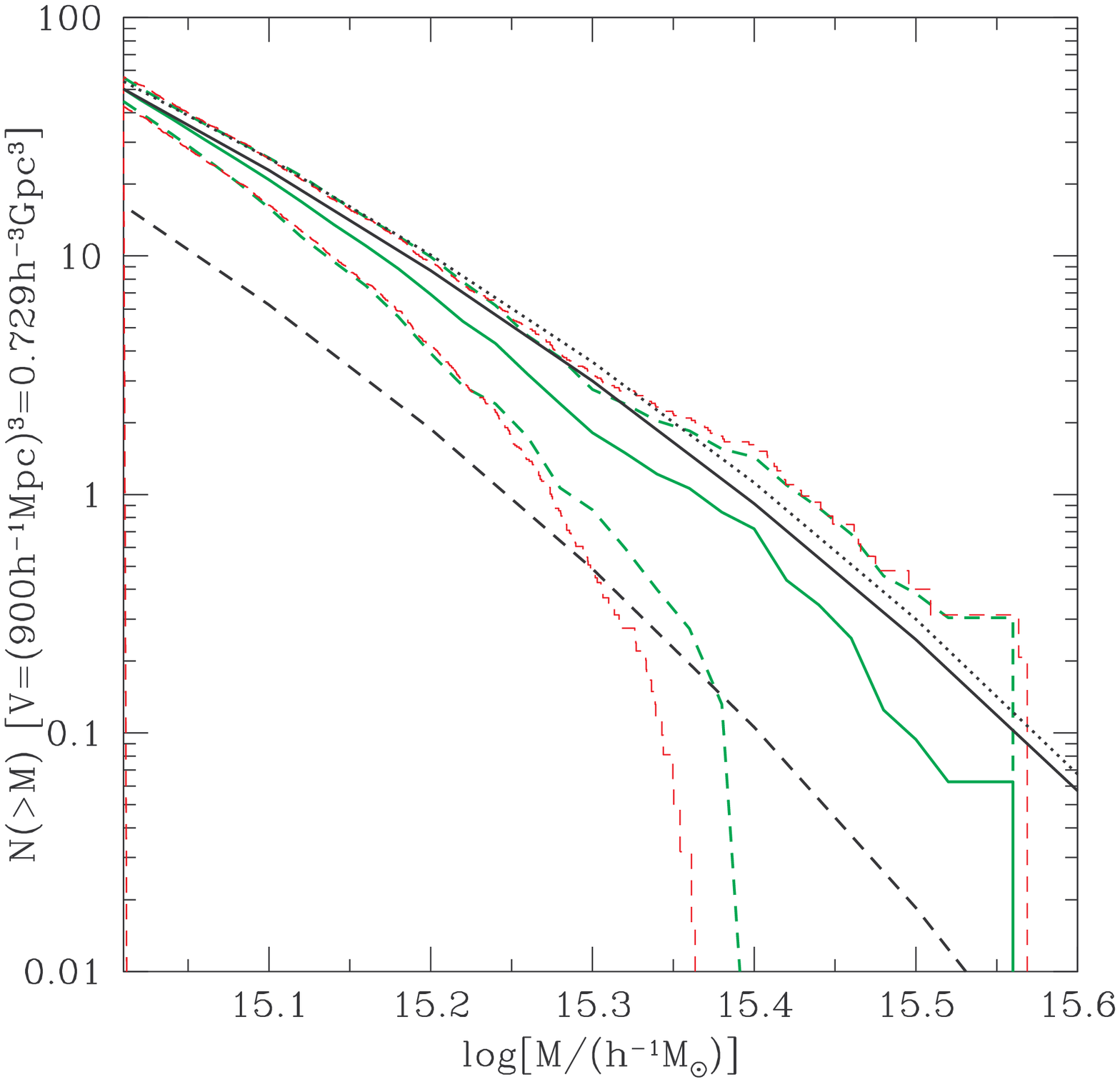}{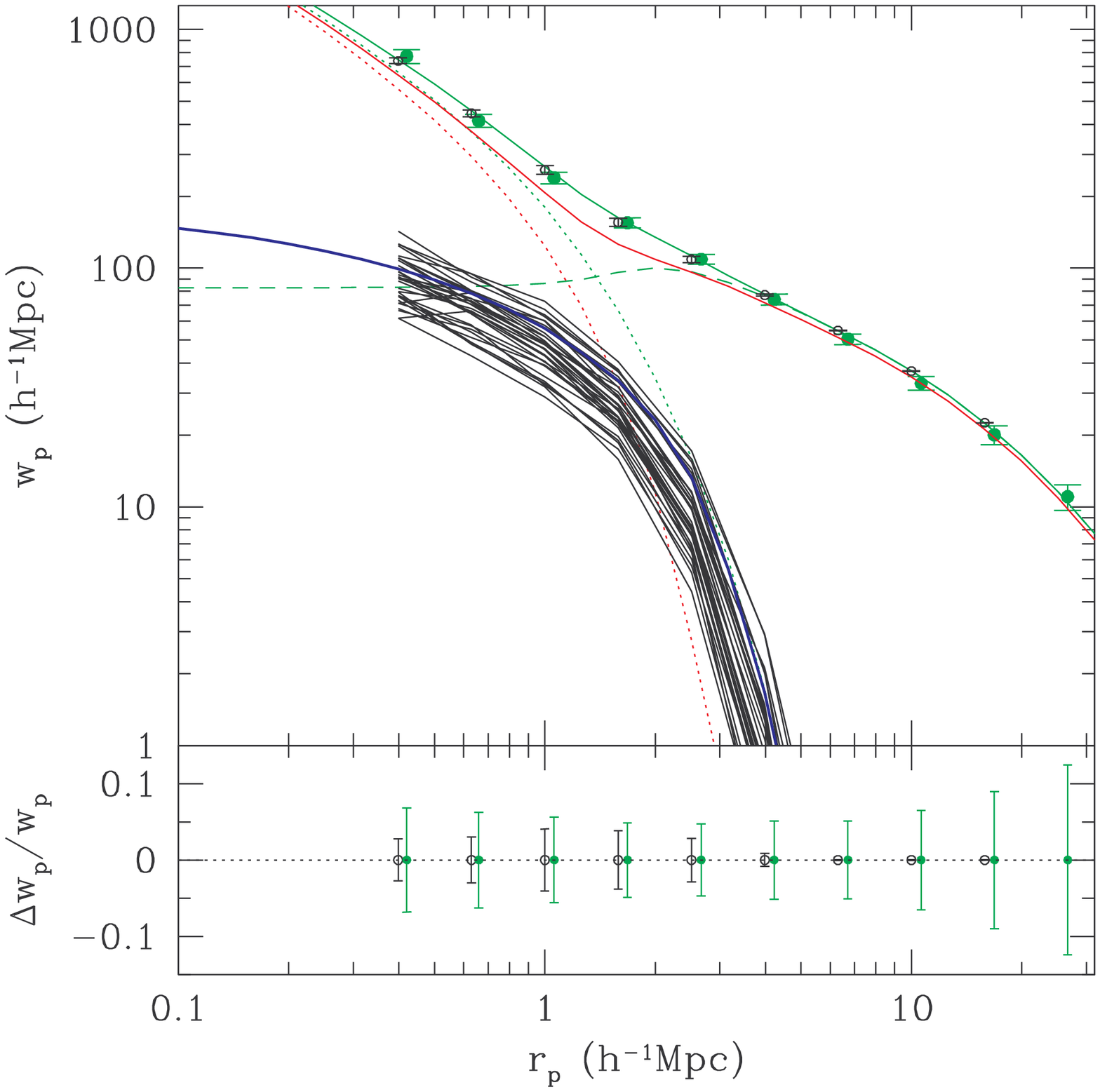}
\caption[]{
{\it (Left):} Halo mass function and its scatter from the 
\citet{Jing07} simulation, for a volume similar in size to the LRG survey 
volume. The green solid curve is the mean halo mass function in the
simulation, the green dashed curves mark the scatter and the red dashed 
curves denote a Poisson scatter around the mean. The black solid, dotted, 
and dashed curves are the Jenkins, Sheth-Tormen, and Press-Schechter mass 
functions, respectively (see the text).
{\it (Right):} Contribution of LRGs in massive halos to small-scale 
$w_p$ and the variation due to the fluctuation of halo mass function in a 
finite volume. The green solid curve is the best fit from a five-parameter 
HOD model to the measurements (green points). The green dotted and dashed 
curves are the one-halo and two-halo terms, respectively. The red solid 
curve is $w_p$ calculated by excluding halos more massive than $10^{15}\hMsun$.
The dotted red curve shows the one-halo term from halos below 
$10^{15}\hMsun$, and the blue curve shows it for halos above this value.
The black solid curves are the one-halo term 
contributions from halos more massive than $10^{15}\hMsun$ in each of the
32 simulation subvolumes that have the same size as the LRG survey volume.
The bottom panel shows the comparison of the diagonal jackknife errors and 
uncertainties introduced by the fluctuation of the number of 
massive halos in the LRG survey volume. See text for more details.
}
\end{figure}

In our HOD modeling of the LRG two-point correlation functions, we 
use a theoretical halo mass function \citep{Jenkins01}, which is a fitting
formula based on $N$-body simulations. In the LRG survey volume,
which is 0.72$h^3{\rm Gpc}^{-3}$ \citep{Eisenstein05b}, fluctuations in the
number of massive halos (e.g., with mass above $10^{15}\hMsun$) are expected.
Since LRGs reside in massive halos, the details of their small scale 
clustering may be sensitive to such fluctuations. In this Appendix, we 
investigate whether these fluctuations introduce any systematic effect on 
the HOD modeling.

\citet{Jing07} performed $N$-body simulations with $1024^3$ particles in
a box of 1800$\hMpc$ on a side. For our investigation, we make use of 
the catalog of massive halos ($> 10^{15}\hMsun$) identified in the 
$z=0.274$ outputs from four independent simulations. The volume of 
each realization is divided into eight octants. Each octant has a volume 
similar to the LRG survey volume, so altogether we have 32 sub-volumes to 
investigate the mass function variations.

The left panel of Figure~\ref{fig:mf_fluctuation} shows the fluctuation of
the number of massive halos among the 32 sub-volumes. The green solid curve
is the mean halo mass function in a sub-volume and the green dashed curves 
mark the scatter around the mean. The scatter turns out to closely follow 
that of a Poisson distribution (the two dashed red curves). 
The fluctuation in the number of halos is about 15\% for $M>10^{15}\hMsun$ 
and increases with halo mass. For comparison, we also plot three 
frequently used theoretical functions for the same cosmological model 
assumed in the simulation. The Jenkins mass function \citep{Jenkins01}, shown 
as the black solid curve, gives a good description of the mean halo mass 
function at $10^{15}\hMsun$ and becomes a factor of two higher at 
$2\times 10^{15}\hMsun$. The Sheth-Tormen mass function (\citealt{Sheth99}; 
black dotted curve) is slightly higher than the Jenkins mass function, while 
the Press-Schechter mass function (\citealt{Press74}; black dashed curve) 
underestimates the mass function by a factor of five in the mass range 
considered here.

The right panel of Figure~\ref{fig:mf_fluctuation} shows the effect of the
fluctuation in the number of massive halos on the small scale clustering
of the $M_g<-21.2$ LRGs. The green solid curve is the best fit to the
measured $w_p$ (green points) from the five-parameter HOD model 
(Appendix~\ref{sec:appendixB}) with the same cosmological model as used 
in the simulation. The green dotted and dashed 
curves are the contributions from the one-halo and two-halo term, 
respectively. With the best-fit HOD model fixed, the red solid curve
show the predicted $w_p$ when only keeping halos less massive than
$10^{15}\hMsun$ in the calculation. The red dotted curve is the
one-halo term from halos less massive than $10^{15}\hMsun$.

The blue solid curve is the mean contribution to the one-halo term from 
halos of $M>10^{15}\hMsun$. The fluctuation in the number of massive
halos would lead to a fluctuation around this mean contribution. To
see this, we populate the massive halos in the 32 sub-volumes according
to the best-fit HOD model and measure the one-halo $w_p$ in each sub-volume.
Black solid curves show the measurements for individual sub-volumes.
Such a fluctuation in the one-halo term introduces uncertainties in
the small scale clustering. By adding the one-halo term from halos
with $M<10^{15}\hMsun$ (red dotted), that from halos above $10^{15}\hMsun$
in each sub-volume (black solid) and the two halo-term (green dashed),
we obtain the black points with error bars reflecting the fluctuation in
the number of massive halos. On small scales, these error bars appear to
have similar (somewhat smaller) amplitude to those from the diagonal elements 
of the covariance matrix (see the comparison in the lower panel), which 
is computed through jackknife technique.

We see that the fluctuation of the number of massive halos in the LRG survey 
volume can be large, e.g., $\sim$15\% for $M>10^{15}\hMsun$ halos and 
$\sim$50\% for $M>2\times10^{15}\hMsun$ halos. This introduces a
fluctuation in the small scale clustering of LRGs. However, our 
investigation suggests that such a fluctuation in $w_p$ should be correctly
reflected in the jackknife covariance matrix. Therefore, it is sufficient 
to use the mean mass function in modeling the two-point correlation functions
of LRGs.

{}

\end{document}